\documentclass[journal,comsoc]{IEEEtran}

\usepackage[T1]{fontenc}% optional T1 font encoding

\usepackage{cite}
\ifCLASSINFOpdf
  \usepackage[pdftex]{graphicx}
\else
\fi
\usepackage[cmex10]{amsmath}

\newcommand{\vect}[1]{\boldsymbol{\mathrm{#1}}}

\usepackage{mathtools}

\usepackage[dvipsnames]{xcolor}
\usepackage{ifpdf}

\usepackage[outdir=./]{epstopdf}

\usepackage[cmintegrals]{newtxmath}
\usepackage{bm}
\usepackage{array}
\ifCLASSOPTIONcompsoc
  \usepackage[caption=false,font=normalsize,labelfont=sf,textfont=sf]{subfig}
\else
  \usepackage[caption=false,font=footnotesize]{subfig}
\fi
\begin{document}
%
% paper title
% Titles are generally capitalized except for words such as a, an, and, as,
% at, but, by, for, in, nor, of, on, or, the, to and up, which are usually
% not capitalized unless they are the first or last word of the title.
% Linebreaks \\ can be used within to get better formatting as desired.
% Do not put math or special symbols in the title.
\title{Experimental Investigation of Frequency Domain Channel Extrapolation in Massive MIMO Systems for Zero-Feedback FDD} %\\ IEEE Communications Society Journals}
%
%
% author names and IEEE memberships
% note positions of commas and nonbreaking spaces ( ~ ) LaTeX will not break
% a structure at a ~ so this keeps an author's name from being broken across
% two lines.
% use \thanks{} to gain access to the first footnote area
% a separate \thanks must be used for each paragraph as LaTeX2e's \thanks
% was not built to handle multiple paragraphs
%

\author{Thomas~Choi,~\IEEEmembership{Student~Member,~IEEE,}
        Fran\c{c}ois~Rottenberg,~\IEEEmembership{Member,~IEEE,}
        Jorge~Gomez-Ponce,~\IEEEmembership{Student~Member,~IEEE,}
        Akshay~Ramesh,~\IEEEmembership{Student~Member,~IEEE,}
        Peng~Luo,~\IEEEmembership{Student~Member,~IEEE,}
        Jianzhong~Zhang,~\IEEEmembership{Fellow,~IEEE,}
        and~Andreas~F.~Molisch,~\IEEEmembership{Fellow,~IEEE}% <-this % stops a space
\thanks{The work was supported by the National Science Foundation (ECCS-1731694), the Belgian National Science Foundation (FRS-FNRS), the Belgian American Educational Foundation (BAEF), the Foreign Fulbright Ecuador SENESCYT Program, and Samsung Research America. A part of this article was presented at the VTC2019-Fall \cite{choi2019channel}.}
\thanks{T. Choi, J. Gomez-Ponce, A. Ramesh, P. Luo, and A. F. Molisch are with the Ming Hsieh Department of Electrical and Computer Engineering, University of Southern California, Los Angeles, CA, USA. J. Gomez-Ponce is also with the ESPOL Polytechnic University, Escuela Superior Politécnica del Litoral, ESPOL, Facultad de Ingenier\'ia en Electricidad y Computaci\'on, Km 30.5 vía Perimetral, P. O. Box 09-01-5863, Guayaquil, Ecuador. F. Rottenberg is with the Université catholique de Louvain, Louvain-la-Neuve, Belgium and the Université libre de Bruxelles, Brussels, Belgium. J. Zhang is with Samsung Research America, Richardson, TX, USA. Corresponding author: Thomas Choi (choit@usc.edu).}% <-this % stops a space
}% <-this % stops a space
\maketitle

% As a general rule, do not put math, special symbols or citations
% in the abstract or keywords.

% Where do we put acknowledgement of Samsung and NSF?
\begin{abstract}
Estimating downlink (DL) channel state information (CSI) in frequency division duplex (FDD) massive multi-input multi-output (MIMO) systems generally requires downlink pilots and feedback overheads. Accordingly, this paper investigates the feasibility of zero-feedback FDD massive MIMO systems based on channel extrapolation. We use the high-resolution parameter estimation (HRPE), specifically the space-alternating generalized expectation-maximization (SAGE) algorithm, to extrapolate the DL CSI based on the extracted parameters of multipath components in the uplink channel. We apply the HRPE to two different channel models: the vector spatial signature (VSS) model and the direction of arrival (DOA) model. We verify these methods through real-world channel data acquired from channel measurement campaigns with two different types of channel sounders: a) a switched array-based, real-time, time-domain, outdoors setup at $3.5~\mathrm{GHz}$, and b) a virtual array-based, high-accuracy, frequency-domain, indoors setup at $2.4$ and $5-7~\mathrm{GHz}$. The performance metrics of the extrapolated channels that we evaluate include the mean squared error, beamforming efficiency, and spectral efficiency in multiuser MIMO scenarios. The results show that the HRPE-based channel extrapolation performs best under the simple VSS model, which does not require array calibration, and if the BS is in an open outdoor environment having line-of-sight (LOS) paths to well-separated users.
\end{abstract}

% Note that keywords are not normally used for peerreview papers.

% For peer review papers, you can put extra information on the cover
% page as needed:
% \ifCLASSOPTIONpeerreview
% \begin{center} \bfseries EDICS Category: 3-BBND \end{center}
% \fi
%
% For peerreview papers, this IEEEtran command inserts a page break and
% creates the second title. It will be ignored for other modes.
\IEEEpeerreviewmaketitle

\section{Introduction}

% The very first letter is a 2 line initial drop letter followed
% by the rest of the first word in caps.
% 
% form to use if the first word consists of a single letter:
% \IEEEPARstart{A}{demo} file is ....
% 
% form to use if you need the single drop letter followed by
% normal text (unknown if ever used by the IEEE):
% \IEEEPARstart{A}{}demo file is ....
% 
% Some journals put the first two words in caps:
% \IEEEPARstart{T}{his demo} file is ....
% 
% Here we have the typical use of a "T" for an initial drop letter
% and "HIS" in caps to complete the first word.
\subsection{Motivation and Problem Statement}
\IEEEPARstart{M}assive~multi-input multi-output (MIMO) systems utilize tens to hundreds of antennas at the base station (BS) to increase the spectral and energy efficiency of wireless networks \cite{marzetta2010noncooperative, marzetta2016fundamentals, bjornson2017massive}. This makes them a promising solution to the rapidly growing number of wireless devices and soaring data capacity requirements. Massive MIMO systems are generally assumed to operate in time division duplex (TDD) mode, where both the uplink (UL) and the downlink (DL) share the same frequency band \cite{bjornson2017massive, bjornson2016massive, flordelis2018tdd}.\footnote{In fact, several researchers have defined that massive MIMO systems operate in TDD mode by default \cite{bjornson2017massive}.} In TDD mode, the massive MIMO system exploits the channel reciprocity, attaining the DL channel state information (CSI) from UL pilots.\footnote{The reciprocity assumption holds only when the transceivers are calibrated \cite{rogalin2014scalable, vieira2017reciprocity, jiang2018framework} and when both the UL and the DL occur within the channel coherence time---usually in scales of milliseconds.}

In comparison, the UL and the DL bands are separated in the frequency division duplex (FDD) mode. Because the channel coherence bandwidth (BW) is almost always much smaller than the duplex spacing \cite{3gpp.36.101}, the channel reciprocity cannot be directly exploited. Therefore, extra overheads, such as DL pilots and feedback, are necessary to attain DL CSI. Also, these overheads scale with the number of antennas at the BS rather than with the total number of antennas at the user equipments (UEs), which may become prohibitive as massive MIMO systems usually have a large number of BS antennas.

Although such high resource demands from the FDD massive MIMO suggest employing the optimal TDD operation for massive MIMO systems, many legacy wireless networks still operate in the FDD frequency spectrum \cite{3gpp.36.101}. Therefore, FDD massive MIMO systems can reduce costs which may arise from modifying the hardware, frequency allocations, and network operation when upgrading the legacy BSs into massive MIMO BSs. One possible solution to enable FDD massive MIMO based on the UL pilots only (like the TDD massive MIMO) is through channel extrapolation (see sec. \ref{literature} for the literature review). 

\begin{figure}[t]
    \centering
    \includegraphics[width=1\linewidth]{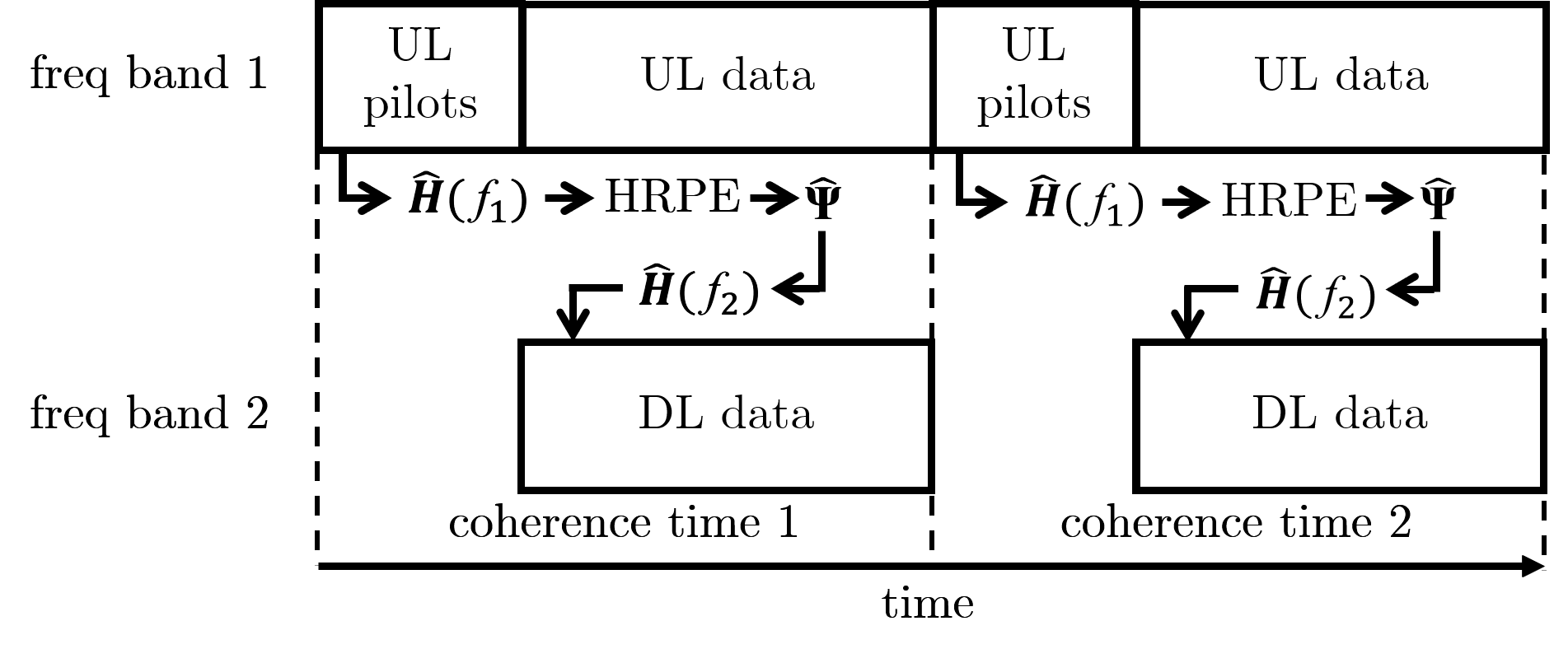}
    \caption{The proposed zero-feedback FDD massive MIMO operation}
    \label{fig:HRPE_extrapol}
\end{figure}

In \cite{rottenberg2019channel, rottenberg2020performance}, we theoretically investigated an approach in which the channel extrapolation uses high-resolution parameter estimation (HRPE). The HRPE determines the parameters of each multipath component (MPC) (e.g., complex amplitude, delay, azimuth direction of arrival (DOA), elevation DOA, etc.) in the UL channel (also known as the training band in the channel extrapolation context) based on the UL pilots. The transfer function of a wireless channel is the complex sum of the contributions of the individual MPCs; thus, the BS can estimate the channel in the DL band without any DL pilots or feedback, assuming that the MPC parameters in the DL channel are the same as those in the UL channel (see Fig. \ref{fig:HRPE_extrapol}). 

The key question we wish to explore is: given a particular UL band and an environment, what is the dependence between the duplex spacing and the extrapolation error? In other words, we explore how far a DL band can be located from a UL band while allowing a reasonable extrapolation. This is fundamentally important, regardless whether a particular duplex spacing is used in a particular system or not, because it can provide insights into the design of future systems.

\subsection{Contributions}    
\begin{enumerate}
    \item In this paper, we apply the framework of \cite{rottenberg2020performance} to real-world channel data acquired through extensive channel measurement campaigns. The measurements were conducted using two different types of massive MIMO channel sounders: a) a switched array-based, real-time, time-domain, outdoors setup at $3.325-3.675$ GHz, and b) a virtual array-based, high-accuracy, frequency-domain, indoors setup at $2.4-2.5$ and $5-7$ GHz.\footnote{Although the selected frequencies of the channel measurements are the frequencies for the TDD operation, the consistency of the results supports the hypothesis that the results can also be applied to other FDD bands.} We chose multiple setups to show that the results are consistent and can be generalized independently of the setup and the environment.
    
    \item We used the space-alternating generalized expectation-maximization (SAGE) algorithm for the HRPE, in two variations: the vector spatial signature (VSS) model \cite{larsen2009performance} and the DOA model \cite{fleury1999channel}. These two models form separate sets of MPC parameters and verify whether or not certain parameters are better suited for the channel extrapolation.

    \item We use three different metrics to assess the evaluated ``accuracy'' and the expected ``performance'' of the extrapolated DL CSI with respect to those of the ground-truth DL CSI. These metrics include the mean squared error (MSE), beamforming efficiency, and spectral efficiency in multiuser MIMO scenarios. Overall, all these metrics investigate the feasibility of zero-feedback FDD massive MIMO systems using HRPE-based channel extrapolation. 
\end{enumerate} 

There are several comments regarding the methodologies and the assumptions of this study:
\begin{enumerate}
    \item Although our channel measurements are conducted in one direction only (i.e., from the UE to the BS), the direction of the measurements {\em does not} impact the results of our investigation. It follows from the fundamental electromagnetic laws that the propagation channel itself is reciprocal at the same frequency within the channel coherence time. Thus, measuring the propagation channel in one direction at the UL and the DL frequencies is equivalent to measuring at the UL frequency from the UE to the BS, and at the DL frequency from the BS to the UE. The same holds for the antenna --- the properties of the antenna depend only on the frequency and {\em not} on whether they are operating as the TX or the RX antenna. The only part of the transmission chain that could be nonreciprocal is the up/down-conversion hardware. This is not an issue because the transceivers are fully calibrated in our channel sounders. %We emphasize while all the channel measurement campaigns were conducted in one direction, from the single-antenna UE to the multi-antenna BS, in a single-input multi-output (SIMO) manner, we don't treat the measured frequency band as the UL band. In fact, due to channel reciprocity, the channel measurement direction does not matter and the UL and the DL bands are selected arbitrarily as separate subsets of the measured frequency band for evaluations. The ``measured CSI at the selected DL band'' is not used during the channel extrapolation and only serves as a ground-truth reference to be compared with the ``extrapolated CSI at the DL band'' constructed using the MPCs extracted at the measured UL band (the training band) $^{(rev1-cmt2)\&(rev1-cmt9)}$.}
    
    \item The actual implementation of the proposed channel extrapolation method for the actual FDD massive MIMO systems can be challenging due to the computational complexity of the HRPE. Instead, our emphasis in this paper is the performance evaluation and the feasibility study of the high-accuracy algorithms for the channel extrapolation. There are several ways to speed-up the MPC extraction using the HRPE, albeit the accuracy will be traded off when the total number of MPCs, number of iterations, parameter resolutions, etc. are reduced. However, such trade-offs are not studied in this paper.
    
    \item For multiuser MIMO studies, we postulate that multiple channel measurements from one UE to a BS at different times is equivalent to measuring multiple UEs to a BS at the same time. This is true only if the channel is completely static. Although we tried to minimize the effects of moving environmental objects, assessing the residual effects quantitatively was not possible.
\end{enumerate}

\subsection{Literature Review} \label{literature}
\subsubsection{Theoretical studies of the FDD massive MIMO}
Several works have studied the feasibility of FDD massive MIMO with reduced overheads. Among the suggested approaches are the compressive sensing (CS) \cite{rao2014distributed}, the long-term channel statistics and the previous signals in a closed-loop manner \cite{choi2014downlink}, or a combination thereof \cite{gao2015spatially}. Other methods utilize the spatial correlation between multiple UEs \cite{adhikary2013joint, jiang2015achievable}, antenna grouping \cite{lee2014antenna}, dominant eigendirections \cite{lee2016exploiting}, nonorthogonal multiple access scheme \cite{ding2016design}, spatial basis expansion model based on the array theory \cite{xie2017unified}, reciprocity based on reverse training \cite{liang2018fdd}, small number of dominant DOAs \cite{shen2018channel}, and the deep learning approach\cite{alrabeiah2019deep, liao2019csi, yang2019deep}.

Similar to this paper, a number of studies have used HRPE to solve the overhead problem. One study used the MUSIC and the ESPIRIT to model channels in disjoint frequency bands \cite{pun2011super}. In \cite{yang2018super} and \cite{urgurlu2016multipath}, the HRPE was proved to be more suited than the CS in exploiting the channel reciprocity in the frequency domain. However, these papers did not verify the proposed methods empirically using real-world channel data.

\subsubsection{Empirical Studies on FDD massive MIMO}
Several measurement-based studies have analyzed the performance of the FDD massive MIMO systems (summarized in Table \ref{table_prev}). Ref.~\cite{flordelis2018tdd} showed that the reciprocity-based TDD massive MIMO performs better than the feedback-based FDD massive MIMO with predetermined grid of beams, especially in the non-line-of-sight (NLOS) cases, based on channel measurements at 2.6 GHz carrier frequency and 50 MHz measurement bandwidth. In \cite{zhang2018directional}, the authors used the DL training and the feedback only toward the four dominant DOAs in order to reduce the overheads. The results showed that spectral efficiency improved by 150 percent compared to that of the full training and the feedback. Channel measurements were then conducted using 64 antennas at 2.4 GHz carrier frequency, 20 MHz training band, and 72 MHz duplex spacing.

Several papers experimentally investigated the ``zero feedback methods'' for FDD massive MIMO. Ref.~\cite{jalden2012channel} used $8 \times 8$ MIMO measurements and a modified maximum likelihood estimator to investigate the extrapolation performance in the spatial and the frequency domains using 5 MHz training band within 2.4--2.45 GHz. Ref.~\cite{vasisht2016eliminating} employed the ``R2-F2'' method, which utilizes the phase changes from inter-antenna separation at the BS to estimate the channel at another frequency band. The proposed FDD system with the extrapolated channel showed high beamforming efficiency. However, only five antennas were used at the BS, with 10 MHz training band within 640--690 MHz. Another zero-feedback method relied on deep learning, which uses a large amount of training data to predict the DL CSI based on the UL CSI \cite{arnold2019enabling}. The measurement setup used 32 antennas at the BS, with 20 MHz training band and 25 MHz duplex spacing within 1.2--1.3 GHz.

The current paper differs from other papers and expands our previous results \cite{choi2019channel, rottenberg2019channel, rottenberg2020performance} by: 1) using another channel model (VSS) that has the advantage of not requiring an array pattern calibration in the HRPE evaluation and channel extrapolation, 2) using an additional channel sounder (virtual array-based, high-accuracy, frequency-domain, indoors setups, at $2.4-2.5$ and at $5-7~\mathrm{GHz}$) to verify consistency of the results in different settings, and 3) employing an additional figure of merit (the spectral efficiency) in the multiuser massive MIMO systems. %The measurement setups include 64 antenna elements cylindrical switched array at 3.325-3.675 GHz using 35 MHz training band \cite{choi2019channel}, 90 positions virtual array at 2.4-2.5 GHz with 20 MHz training band, and 72 positions virtual array at 5-7 GHz with 50 MHz training band. 
More recently, \cite{hong2020fdd} used a similar channel inference method using the SAGE and four other different types of calibration methods. However, only a DOA model that excludes the elevation DOA was used in the paper. Moreover, the measurement BW was smaller, at less locations, and the multiuser scenarios were not considered.
\begin{center}
\begin{table}[!t]
\centering
% increase table row spacing, adjust to taste
\renewcommand{\arraystretch}{1}
%if using array.sty, it might be a good idea to tweak the value of \extrarowheight as needed to properly center the text within the cells
\caption{List of empirical FDD massive MIMO research}
\label{table_prev}
\centering
% Some packages, such as MDW tools, offer better commands for aking tables
% than the plain LaTeX2e tabular which is used here.
\begin{tabular}{|c | m{4.9cm}| c| }

\hline
Institution & DL CSI selection/estimation method & Feedback\\
\hline
\hline
Lund \cite{flordelis2018tdd} & the best beam among a grid of beams & yes \\
\hline
Rice \cite{zhang2018directional} & maximum likelihood method \cite{krim1996two} for the dominant DOAs & yes\\
\hline
Ericsson \cite{jalden2012channel} & modified maximum likelihood method \cite{medbo2012efficiency} & no\\
\hline
MIT \cite{vasisht2016eliminating} & R2-F2 \cite{vasisht2016eliminating} & no\\
\hline
Stuttgart \cite{arnold2019enabling} & deep learning & no\\
\hline
USC & SAGE VSS \cite{larsen2009performance} / DOA \cite{fleury1999channel} & no\\
\hline
\end{tabular}
\end{table}
\end{center}
\section{VSS and DOA Channel Models} \label{model}
\begin{figure*}
    \centering
    \includegraphics[width=0.75\linewidth]{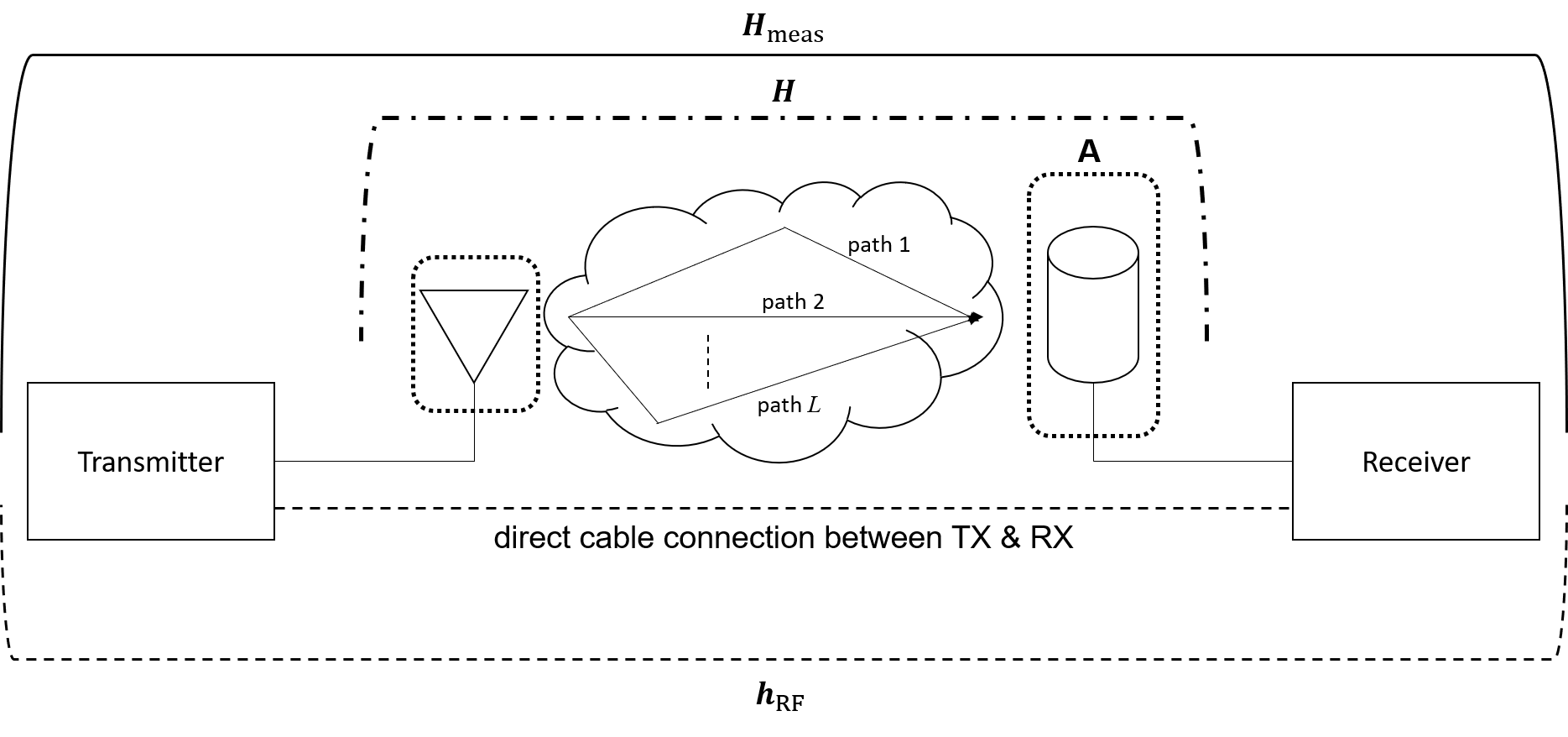}
    \includegraphics[width=0.75\linewidth]{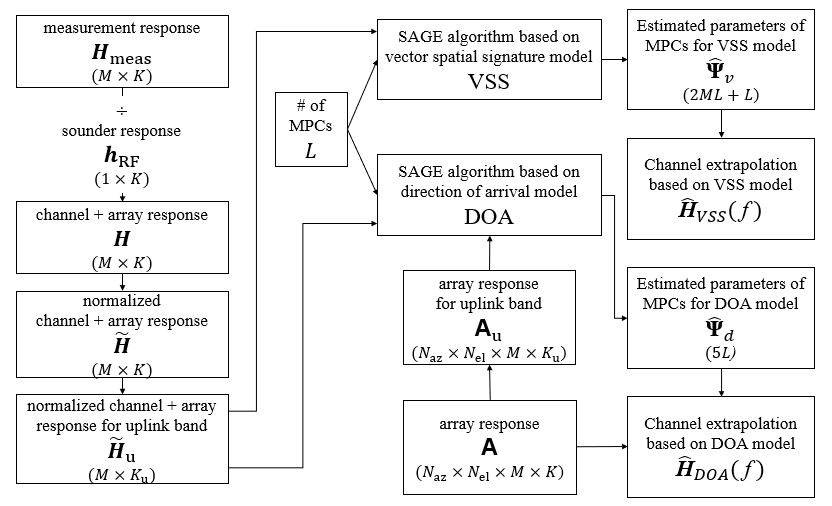}
    \caption{A detailed process of the channel extrapolation using the SAGE algorithm based on the VSS and the DOA models}
    \label{fig:VSS_DOA}
\end{figure*}

In this work, we 1) estimate the MPC parameters using an HRPE based on the parametric channel models, and subsequently get complete descriptions of the channels at the training (UL) band (an arbitrarily selected subset of the measured frequency band); and 2) extrapolate the estimated CSI to the other (DL) frequency band (the complement band of the training band within the measured frequency band) using the parameters obtained from the training band. The overall inputs and outputs of the HRPEs depending on two different parametric channel models are shown in Fig. \ref{fig:VSS_DOA}. These will be explained in the succeeding subsections. 

\subsection{Channel Matrix for the HRPE}
 First, we select a raw channel matrix measured by the receiver of a channel sounder, $\bm{H}_{\rm{meas}}$, with dimension $M \times K$. $M$ is the number of antennas at the BS, whereas $K$ is the number of frequency samples of the measurement. The subscript $meas$ indicates that the data have not yet been compensated for the response of the calibrated radio frequency (RF) system, $\bm{h}_{\rm{RF}}$. $\bm{h}_{\rm{RF}} = [h_{\rm{RF},\it{1}},h_{\rm{RF},\it{2}},\dots,h_{\rm{RF},\it{K}}]^\mathsf{T}$ is a $K\times1$ vector of scalars, which is the back-to-back calibration response of the sounder transmitter (TX) and the receiver (RX) connected by a cable (excluding the antennas and the channel).\footnote{$\bm{h}_{\rm{RF}}$ is a 1-D frequency response because each of the switched and the virtual array setups we used relies on a single RF chain for all antennas---it can be a matrix for other setups if multiple RF chains are used. This notation must not be confused with the impulse response notation, which usually employs lower case letter.} The compensated frequency response of the channel and the antennas only, $\bm{H}$, is attained by

\begin{equation}
\bm{H} = \bm{H}_{\rm{meas}} \cdot \begin{bmatrix}
{(h_{\rm{RF},\it{1}})}^{-1}&0&\cdots &0 \\
0&{(h_{\rm{RF},\it{2}})}^{-1}&\cdots &0 \\
\vdots & \vdots & \ddots & \vdots\\
0&0&\cdots &{(h_{\rm{RF},\it{K}})}^{-1}
\end{bmatrix}.
\end{equation}
%where ${h_{\rm{RF},\it{k}}}^{-1}$ represents the inverse of the RF system frequency response at $k$th frequency index. 

We define averaged power of $\bm{H}$ as ${\mu_{\bm{H}}}^2 = \frac{||\bm{H}||_{F}^2}{MK}$, where $||\cdot||_F$ is the Frobenius norm, and the normalized channel measurement matrix as $\tilde{\bm{H}}=\frac{\bm{H}}{\mu_{\bm{H}}}$. If $K_{\rm{u}}\leq K$ is the number of frequency points in the arbitrarily selected training band within the measurement band, an $M \times K_{\rm{u}}$ subset of $\tilde{\bm{H}}$, $\tilde{\bm{H}}_{\rm{u}} = \tilde{\bm{H}}_{(:,~a:a+K_{\rm{u}}-1)}$ with $1\leq a \leq K-K_{\rm{u}}+1$, represents the measured channel matrix at the training band. $\tilde{\bm{H}}_{\rm{u}}$ becomes the input for the HRPE. 

Although there are many well-known HRPEs (e.g., MUSIC \cite{schmidt1986multiple}, ESPRIT \cite{roy1989esprit}, CLEAN \cite{hogbom1974aperture}, and RiMAX \cite{richter2005estimation}), we selected the SAGE algorithm arbitrarily, as the extrapolation results during the preliminary analysis were dependent on the channel models rather than on the type of HRPEs. Two channel models (i.e., the VSS and the DOA) are used to attain the different types and values of the MPC parameters. This will be further discussed in the succeeding subsections. A more detailed description of the SAGE algorithm is given in Appendix \ref{appendix}.

\subsection{Vector Spatial Signature (VSS) Model}
The main difference between the VSS and the DOA models is that the former does not require a calibrated array pattern, $\bm{\mathsf{A}}$, as it does not estimate the DOAs of MPCs. The only input for the VSS algorithm, other than the normalized channel measurement matrix at the UL band, $\tilde{\bm{H}}_{\rm{u}}$, is $L$, which is the number of MPCs (Fig. \ref{fig:VSS_DOA}). In terms of channel extrapolation, describing the channel with large $L$ may result in lower prediction accuracy due to overfitting. Furthermore, MPCs with lower amplitudes will be more prone to noise, and thus may deteriorate the overall channel prediction in the extrapolated band. Therefore, the selected $L$ needs to be sufficiently large to estimate the training band accurately (MSE $<-10$ dB), but small enough to prevent overfitting. This compromise was found empirically, and the choice of $L$ accordingly differed between measurements under different environments \cite{choi2019channel}.

\begin{table*}[!t]
% increase table row spacing, adjust to taste
\renewcommand{\arraystretch}{1}
%if using array.sty, it might be a good idea to tweak the value of \extrarowheight as needed to properly center the text within the cells
\caption{Estimated MPC parameters in the VSS and the DOA channel models}
\label{table_example}
\centering
% Some packages, such as MDW tools, offer better commands for aking tables
% than the plain LaTeX2e tabular which is used here.
\begin{tabular}{|c||c| c| c|}
\hline
Model & Estimated MPC parameters & Channel model\\
\hline
VSS & $\hat{\bm{\psi}}_{v} = [\bm{\hat{\mathsf{a}}}_{v}, \hat{\tau}_{v}]$ & $\hat{\bm{H}}_{VSS}(f) = \sum_{v=1}^{L} \bm{\hat{\mathsf{a}}_{v}} e^{-j2\pi f \hat{\tau}_{v}}$\\
\hline
DOA & $\hat{\bm{\psi}}_{d} = [\hat{\alpha}_{d}, \hat{\tau}_{d}, \hat{\phi}_{d}, \hat{\theta}_{d}]$ & $\bm{\hat{H}}_{DOA}(f) = \sum_{d=1}^{L} \hat{\alpha}_{d} \bm{\mathsf{A}}(\hat{\phi}_{d},\hat{\theta}_{d},f)e^{-j2\pi f \hat{\tau}_{d}}$\\
\hline
\end{tabular}
\end{table*}

The first parameter of the VSS model is the estimated delay, $\hat{\tau}_{v}$, where $v$ is the index of the MPC. The second parameter is called the ``vector spatial signature (VSS)'', represented by $\hat{\bm{\mathsf{a}}}_{v}$. $\hat{\bm{\mathsf{a}}}_{v}$ is a \emph{frequency independent} complex vector with size $M \times 1$. According to \cite{larsen2009performance}, the VSSs are ``not explicit functions of DOA, but instead \emph{abstractly} represent the response of the array pattern for the path $v$ with the delay $\tau_{v}$''. 

The extrapolation range is limited in the VSS model because in practice, most array patterns are frequency-dependent. However, estimating within the training band will be very accurate for massive MIMO systems because the number of the estimated parameters in the VSS model scales with $M$; these parameters can then be adjusted to best estimate the channel response. 

The VSS channel model is overall expressed as:

\begin{equation}
\label{vss_eq}
\hat{\bm{H}}_{VSS}(f) = \sum_{v=1}^{L} \bm{\hat{\mathsf{a}}_{v}} e^{-j2\pi f \hat{\tau}_{v}}.
\end{equation}

Once both $\hat{\bm{\mathsf{a}}}_{v}$ and $\hat{\tau}_{v}$ are estimated for all $L$ paths from the training band, the CSI at the desired frequency (such as the DL frequency in the FDD systems), $f_{\rm d}$, can be attained by setting $f=f_{\rm d}$ in eq. (\ref{vss_eq}). The VSS model hence assumes that only the phase changes with frequency.

In total, the VSS model estimates the $2ML+L$ real-valued parameters, where the first term $2ML$ is the total number of real and the imaginary values (2) of the vector spatial signatures for all antennas ($M$) over all paths ($L$), and the second term $L$ is the total number of delays.

\subsection{Direction of Arrival (DOA) Model}
As the name indicates, the DOA model requires a frequency dependent calibrated antenna array pattern, $\bm{\mathsf{A}}$, to determine the DOAs of the incoming MPCs. Aside from the topic of channel extrapolation, this model is useful when the angular information of the channel is needed. Like the VSS model, the DOA model also requires the normalized channel measurement matrix in the UL band, $\tilde{\bm{H}}_{\rm{u}}$, and the total number of MPCs, $L$ (Fig. \ref{fig:VSS_DOA}).

$\bm{\mathsf{A}}$ is attained differently for switched array and virtual array \cite{choi2019channel}. In a switched array, a reference antenna with a known pattern is positioned on one side of the anechoic chamber, and the switched array is positioned on another side that is supported by a rotating positioner. The array rotates to a certain azimuth and elevation position, and then switches on a selected antenna element. A vector network analyzer (VNA) then sweeps across the frequency to record the channel frequency response. Subsequently, the next antenna element is turned on for another VNA sweep. After obtaining the transfer function for each antenna element, the antenna array rotates to a new azimuth and elevation position; the process repeats to record the channel frequency response at the selected position. 

In the end, a 4-D calibration data, with dimension $N_{\rm{az}} \times N_{\rm{el}} \times M \times K$, is created. $N_{\rm{az}}$ and $N_{\rm{el}}$ are the number of azimuth steps and number of elevation steps during calibration, respectively. The calibrated array pattern at the training band, $\bm{\mathsf{A}}_{\rm{u}}$, is selected as an input for the HRPE, with dimension $N_{\rm{az}} \times N_{\rm{el}} \times M \times K_{\rm{u}}$. The complement set of $\bm{\mathsf{A}}_{\rm{u}}$ within $\bm{\mathsf{A}}$ will be used when the channel is extrapolated to the frequency outside the training band.

The virtual array calibration, however, does not rely on switching; thus, a calibration data of a single antenna with dimension $N_{\rm{az}} \times N_{\rm{el}} \times K$ is created first. This 3-D calibration data of a single antenna is numerically rotated or moved $M$ times to form a 4-D calibration data of a virtual antenna array with the selected geometry and numbers. The impact of the rotation/movement is computed geometrically, and not measured explicitly. The formulated calibration data, $\bm{\mathsf{A}}$, with dimension $N_{\rm{az}} \times N_{\rm{el}} \times M \times K$, are then used in the same way as the switched-array calibration data.

We have shown in \cite{choi2019channel} the output parameters of the MPCs attained through the SAGE algorithm based on the DOA model and channel extrapolation results in an anechoic chamber and outdoors. The estimated parameters of each MPC in the DOA model include the complex amplitude, $\hat{\alpha}$, delay, $\hat{\tau}$, azimuth DOA, $\hat{\phi}$, and elevation DOA, $\hat{\theta}$. In this paper, the subscript $d$ is added to the parameters ($\hat{\alpha}_d$, $\hat{\tau}_d$, $\hat{\phi}_d$, and $\hat{\theta}_d$) as an index of the MPC in order to distinguish between the parameters of the VSS and the DOA models. These parameters are summarized in Table \ref{table_example}. There are $5L$ real-valued parameters (the real and the imaginary values of the complex amplitude and other three parameters per path over all paths) to estimate, which are usually much less than $2ML + L$ parameters in the VSS model.

The estimated parameters from the training band are put back into the channel model such that the channel frequency response can be estimated at a selected frequency, $f$. The DOA channel model is as follows:

\begin{equation}
    \bm{\hat{H}}_{DOA}(f) = \sum_{d=1}^{L} \hat{\alpha}_{d} \bm{\mathsf{A}}(\hat{\phi}_{d},\hat{\theta}_{d},f)e^{-j2\pi f \hat{\tau}_{d}}
\end{equation}
where $\bm{\mathsf{A}}(\hat{\phi}_{d},\hat{\theta}_{d},f)$ is the $M \times 1$ 1-D slice of the 4-D array response over $M$ antenna elements (switched array) or positions (virtual array). It is dependent on the estimated azimuth DOA from the training band, estimated elevation DOA from the training band, and frequency of choice. The array pattern at a specific frequency is obtained, either through including the frequency point during calibration or through interpolation if the desired frequency is between two measured frequency points during the calibration. The DOA model assumes both the array response and the phase changes with frequency. 

\section{Performance Metrics}
We select the following three types of performance metrics to assess the performance of the extrapolation: the MSE, beamforming efficiency, and spectral efficiency for multiuser scenarios.

\subsection{Mean Squared Error (MSE)}
The MSE averages the squared magnitude of the differences between the normalized complex channel response of a measured (ground-truth) channel and the estimated channel at the selected frequency over all antennas:

\begin{equation}
    MSE(f) \overset{\Delta}{=}\frac{||\bm{\tilde{H}}(f)-\bm{\hat{H}}(f)||_2^2}{M}
\end{equation}
where $\bm{\hat{H}}(f)$ is either $\bm{\hat{H}}_{DOA}(f)$ or $\bm{\hat{H}}_{VSS}(f)$ with dimension $M \times 1$ and $|| \cdot||_2$ is the Euclidean norm. If $f$ lies within the training band, then the MSE will be a measure of the interpolation performance. In comparison, if $f$ lies outside the training band, then the MSE will be a measure of the extrapolation performance. While the MSE measures the absolute accuracy of the extrapolated channel, it neglects the ability of the UEs to compensate for a common phase and amplitude error. The MSE will be represented on a dB scale. 

\subsection{Beamforming Efficiency}
Massive MIMO array obtains a beamforming gain (BG) by combining constructive contributions of the many antenna elements within the array. One common way to optimize the BG (in a single-user case) is through the maximum-ratio combining (the matched filtering) based on the estimated channel response. The beamforming efficiency (BE) indicates how the BG with the estimated CSI compares with that based on the measured (ground-truth) CSI. The BG with the measured CSI, the estimated CSI, and the uniform beamforming are expressed as follows:

\begin{equation}
    BG_{meas} (f)\overset{\Delta}=||\bm{\tilde{H}}(f)||_2^2
    \label{perfect}
\end{equation}
\begin{equation}
    BG_{est} (f)\overset{\Delta}=\frac{|{\bm{\hat{H}}(f)}^\dagger \bm{\tilde{H}}(f)|^2}{||\bm{\hat{H}}(f)||_2^2}
    \label{estimated}
\end{equation}
\begin{equation}
    BG_{uni} (f)\overset{\Delta}=\frac{||\bm{\tilde{H}}(f)||_2^2}{M}
\end{equation}
where $\dagger$ is the conjugate transpose operator. Therefore, using eq. (\ref{perfect}) and (\ref{estimated}), the BE is expressed as follows:

\begin{equation}
    BE (f)\overset{\Delta}=\frac{BG_{est}(f)}{BG_{meas}(f)}=\frac{|{\bm{\hat{H}}(f)}^\dagger \bm{\tilde{H}}(f)|^2}{||\bm{\hat{H}}(f)||_2^2||\bm{\tilde{H}}(f)||_2^2}.
\end{equation}
This value ranges from 0 to 1, with 1 indicating full efficiency. Similar to the case of the MSE, the BE will be represented on a dB scale. 

\subsection{Spectral Efficiency in Multiuser MIMO Systems}
The earlier metrics are based on the single-user assumption. The spectral efficiency in every UE of a multiuser MIMO system can be determined by single-user measurements at multiple locations. The spectral efficiency of the UE $n$ at frequency $f$ is represented as:

\begin{equation}
    C^{(n)}(f) = \mathrm{log}_2(1+\mathrm{SINR}^{(n)}(f))\;[\mathrm{bits/s/Hz}]
\end{equation}
where $\mathrm{SINR}^{(n)}(f)$ is a signal-to-interference-plus-noise ratio at a given frequency and UE. The index of each UE, $n$, varies from $1$ to $N$. The received signal by the $n$th UE at frequency $f$ during the DL phase is:

\begin{align}
    r^{(n)}(f) &= \bm{H}^{(n)}(f)^{\mathsf{T}} \bm{\tilde{G}}^{(n)}(f)s^{(n)} \notag\\
    &\quad+ \sum_{n' \neq n}^N\bm{H}^{(n)}(f)^{\mathsf{T}}\bm{\tilde{G}}^{(n')}(f)s^{(n')} + w^{(n)}
\end{align}
where $\mathsf{T}$ is a transpose operator; $r^{(n)}(f)$ is a complex value received by the UE $n$ at the selected frequency, $f$; $\bm{H}^{(n)}(f)$ is the ground-truth channel vector for the UE $n$ at frequency $f$ from every antenna, with dimension $M \times 1$; $\bm{\tilde{G}}^{(n)}(f)$ is the normalized precoding vector of the UE $n$ at frequency $f$ from every antenna with dimension $M \times 1$; $s^{(n)}$ is a transmitted signal from the BS for the UE $n$;and $w^{(n)}$ is a noise received by the UE $n$. 

The first term indicates the beamforming signal for the UE $n$, whereas the second term indicates the interference from the signals intended for other UEs received by the UE $n$. Therefore, $\mathrm{SINR}^{(n)}(f)$ is expressed as:

\begin{equation}
    \mathrm{SINR}^{(n)}(f) = \frac{|\bm{H}^{(n)}(f)^{\mathsf{T}}\bm{\tilde{G}}^{(n)}(f)|^2{\sigma_{s^{(n)}}}^{2}}{\sum_{n' \neq n}^{N}|\bm{H}^{(n)}(f)^{\mathsf{T}}\bm{\tilde{G}}^{(n')}(f)|^2 {\sigma_{s^{(n')}}}^{2} + {\sigma_{w^{(n)}}}^{2}}
\end{equation}
where ${\sigma_{s^{(n)}}}^2$ and ${\sigma_{w^{(n)}}}^2$ are the variances of the signal and the noise, assumed to be the same for all $N$ UEs (i.e., all UEs are expected to experience the same transmit power from the BS and the same noise power).

The normalized precoding vector, $\bm{\tilde{G}}^{(n)}(f)$, can be created as follows. First, 
\begin{align}
\begin{split}
\bm{\hat{H}}_{MU}(f) &= [\bm{\hat{H}}^{(1)}(f), \bm{\hat{H}}^{(2)}(f), ... ,\bm{\hat{H}}^{(N)}(f)] \\ &= \begin{bmatrix}
{\hat{H}}^{(1)}_{1}(f)&{\hat{H}}^{(2)}_{1}(f)&\cdots &{\hat{H}}^{(N)}_{1}(f) \\
{\hat{H}}^{(1)}_{2}(f)&{\hat{H}}^{(2)}_{2}(f)&\cdots &{\hat{H}}^{(N)}_{2}(f) \\
\vdots & \vdots & \ddots & \vdots\\
{\hat{H}}^{(1)}_{M}(f)&{\hat{H}}^{(2)}_{M}(f)&\cdots &{\hat{H}}^{(N)}_{M}(f)
\end{bmatrix} \end{split}\end{align} 
where ${\hat{H}}^{(n)}_{m}(f)$ represents an estimated complex channel value between the BS antenna $m$ and the UE $n$ at frequency $f$. Then, the precoding matrix for the $N$ UEs can be determined in two ways: the maximum ratio (MR) and the zero-forcing (ZF):

\begin{equation}
    \bm{G}_{MR}(f) = \bm{\hat{H}}_{MU}(f)^\mathsf{\dagger}
\end{equation}
\begin{equation}
    \bm{G}_{ZF}(f) = \bm{\hat{H}}_{MU}(f)^\mathsf{\dagger}(\bm{\hat{H}}_{MU}(f)\bm{\hat{H}}_{MU}(f)^\mathsf{\dagger})^{-1}.
\end{equation}

Dimension of both matrices is $N \times M$. $\bm{G}^{(n)}(f)$ is a precoding vector of the UE $n$ at frequency $f$ with dimension $M \times 1$ (the transpose of row $n$ of either $\bm{G}_{MR}(f)$ or $\bm{G}_{ZF}(f)$). Finally, we normalize this precoding vector to get the normalized precoding vector per UE, $\bm{\tilde{G}}^{(n)}(f) = \frac{\bm{G}^{(n)}(f)}{||\bm{G}^{(n)}(f)||_2}$. 

\section{Measurement Setups and Settings} \label{measurement_s}
We conducted the measurement campaigns using two types of channel sounders under different environments. Each of these methods has respective advantages and drawbacks. The ``time-domain'' setup, which is based on the arbitrary waveform generator (AWG) and the digitizer combined with the fast-switching array, enables measurements under fast-varying environments. On the other hand, the “frequency-domain” setup, which is based on the vector network analyzer (VNA) combined with a single antenna forming a virtual array, enables high-precision measurements without antenna coupling or element impairments. The list of commercial off-the-shelf hardware used to build the sounders is given in Table \ref{table_equip}.

\begin{table*}[!t]
% increase table row spacing, adjust to taste
\renewcommand{\arraystretch}{1}
%if using array.sty, it might be a good idea to tweak the value of \extrarowheight as needed to properly center the text within the cells
\caption{List of channel sounder equipment}
\label{table_equip}
\centering
% Some packages, such as MDW tools, offer better commands for aking tables
% than the plain LaTeX2e tabular which is used here.
\begin{tabular}{|c | c| c| c| c|}
\hline
Type & Real-time sounder with switched array & VNA-based sounder with rotating horn\\
\hline
\hline
Transmitter & Agilent N8241A AWG & Keysight E5080A VNA \\
\hline
Receiver & NI PXIe-5160 Oscilloscope & Keysight E5080A VNA\\
\hline
Clock & Precision Test Systems GPS10eR & internal clock within VNA\\
\hline
Mixer & Mini-Circuits ZEM-4300MH+ & N/A\\
\hline
Local Oscillator & Phase Matrix FSW-0020 & N/A\\
\hline
Amplifier & \begin{tabular}{@{}c@{}} Mini-Circuits ZFL-500LN\\ Mini-Circuits ZHL-100W-382+\\ Wenteq ABP1500-03-3730 \\ Wenteq ABL0600-33-4009\end{tabular} & \begin{tabular}{@{}c@{}} Wenteq ABP1500-03-3730\\ Wenteq ABL0800-12-3315\end{tabular}\\
\hline
Filters & Pasternack PE8713& \begin{tabular}{@{}c@{}} 
Mini-Circuits ZBP-2450+\\
Mini-Circuits VHF3800\\ 
Mini-Circuits VLF7200+\end{tabular}\\
\hline
Antennas & \begin{tabular}{@{}c@{}} Lab-built stacked patch array\\ Cobham XPO2V-0.3-10.0/1381\end{tabular} & \begin{tabular}{@{}c@{}} L-com HG2420EG-NF\\ A-INFO LB-159-20-C-SF\\ Cobham XPO2V-0.3-10.0/1381\end{tabular}\\
\hline
Switch & \begin{tabular}{@{}c@{}} Pulsar SW8AD\\ Pulsar SW16AD\end{tabular} & N/A\\
\hline
Positioner & N/A & Dams DCP252A \\
\hline
\end{tabular}
\end{table*}

In both channel sounders, an omnidirectional antenna is used at the TX that emulates a UE. On the RX (BS) side, the switched array is used during the outdoor measurements and the rotating horn during the indoor measurements. In order to consider the measured channel as the reference ``ground-truth'' channel such that the estimated channel can be compared with, a high signal-to-noise ratio (SNR) was necessary during the measurement. Therefore, we used an effective isotropic radiated power (EIRP) up to 40 dBm during the outdoor measurements and another EIRP up to 28 dBm during the indoor measurements.

Because the outdoor measurements were conducted without any vehicular movements, the changes in the environment occurred only from walking pedestrians and moving vegetation. Even with a 10 km/h maximum speed assumption, the coherence time is in the order of 30 ms at 3.5 GHz frequency. The switched array-based, time-domain channel sounder captures the channel responses between the UE antenna and all the BS antennas in 10.24 ms, which is well within the outdoor channel coherence time. In the indoor measurement, meanwhile, although the virtual array-based, frequency-domain channel sounder takes several minutes to capture the channel, we conducted the measurements late at night when there were no people. This then ensured that there was no movement in the environment, leading to (theoretically) infinite coherence time.

\subsection{Outdoor Measurements with Switched Array}
\begin{figure}[!t]
    \centering
    \subfloat[Real-time sounder/switched array]{\includegraphics[height=3.3in]{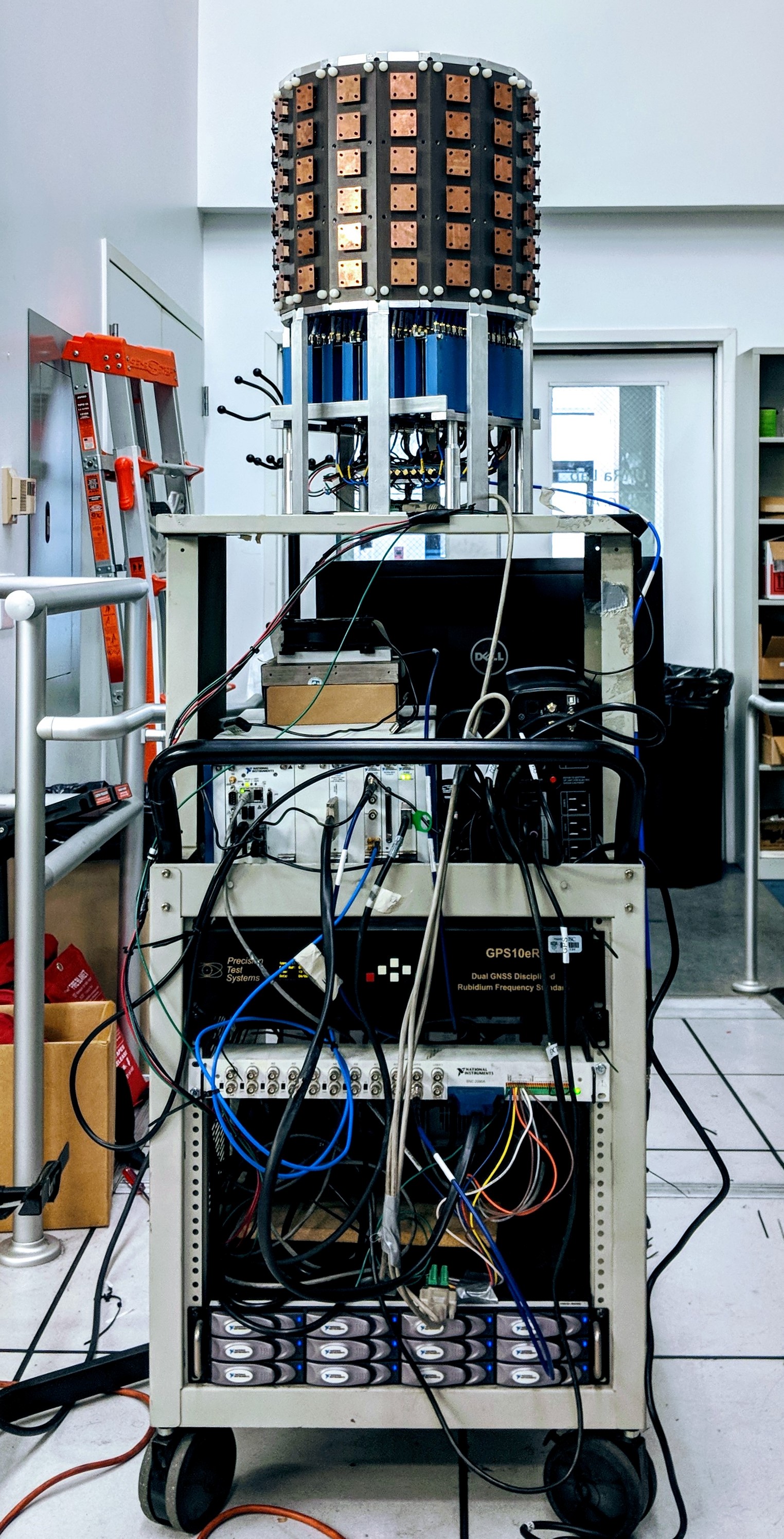}%
    \label{switched_array}}
    \hfil
    \subfloat[VNA-based sounder/rotating horn]{\includegraphics[height=3.3in]{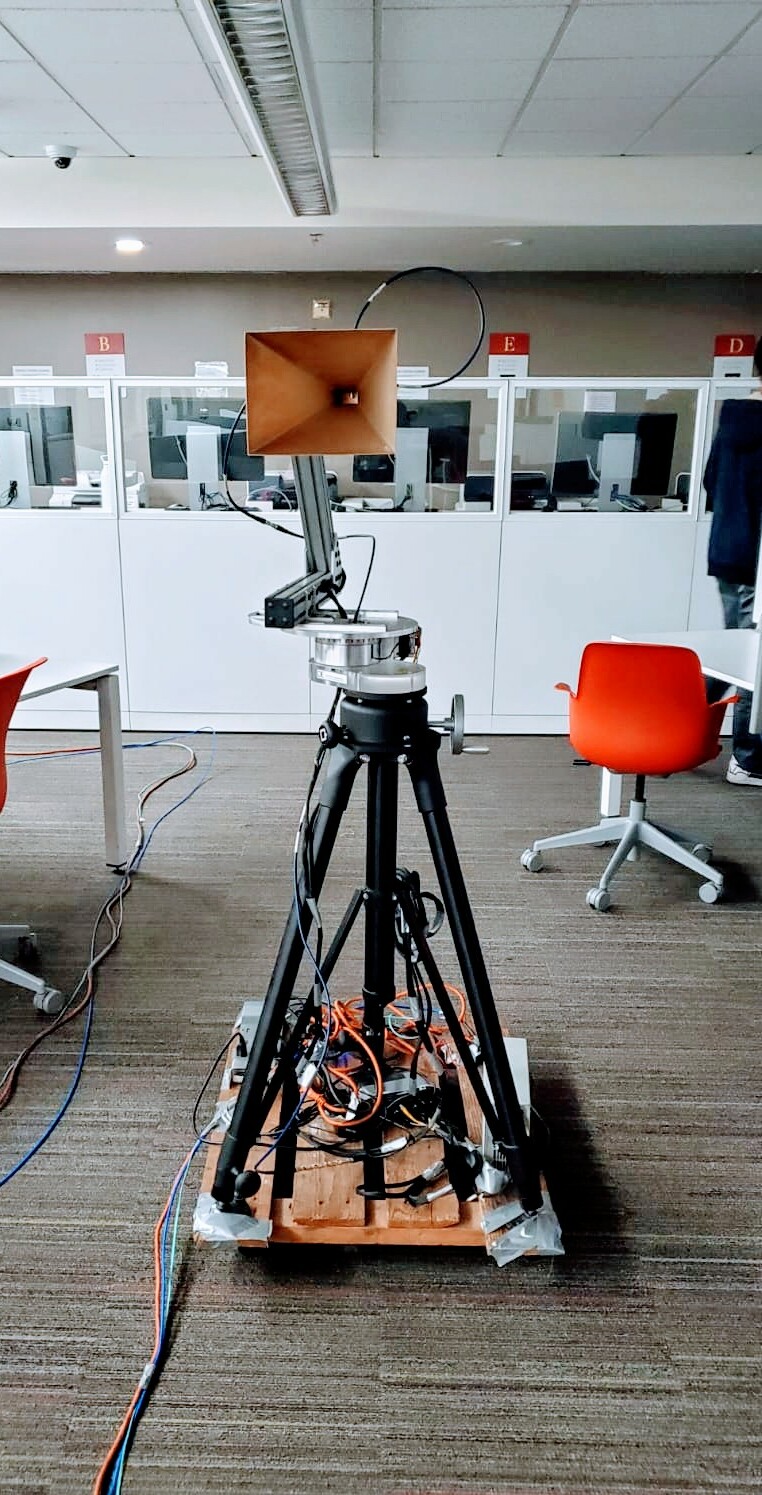}%
    \label{virtual_array}}
    \caption{Two types of channel sounders used for measuring the channel responses of dynamic outdoor and static indoor environments}
    \label{fig_sim}
\end{figure}

To measure the channel characteristics of the outdoor channels with short coherence time, we used a switched array with 64 antenna elements as the RX, thereby emulating a massive MIMO BS (Fig. \ref{switched_array}). The antenna array is cylindrical and has 16 columns of $6\times 1$ linear antenna array. Each antenna has a stacked patch design, which increases the beamwidth and the BW as compared to the conventional patch antennas. Four antenna elements per column are used during the measurement since we use the top and bottom antenna elements (per each column) as the ``dummy'' antenna elements. This then results in 64 active antenna elements. Although each antenna element has two ports (vertical/horizontal polarization) , we consider only the vertically polarized ports because the TX uses a vertically polarized antenna. The radiation patterns of the antenna elements are also described in \cite{choi2019channel}.

These 64 active antenna elements are then connected to eight $16 \times 1$ switches. The switches are cascaded with one $8 \times 1$ switch, thereby resulting in a single RF chain at the RX. The single RF chain simplifies the RF ``back-to-back'' calibration and the sounder operation. The switches are controlled with a digital control interface. %The sounder captures channel responses across all 64 antenna elements in 10.24 ms. This is smaller than the channel coherence time since any possible movement can only come from slow-moving objects such as people and tree leaves. 

The sounder operates in the 3.5 GHz frequency range, with a measurement band ranging from 3.325 to 3.675 GHz. It uses a multitone waveform with low crest factor \cite{friese1997multitone} to achieve low peak to average power ratio, thereby allowing the system to operate close to the 1-dB compression point of the power amplifier. The subcarrier spacing is 125 kHz, resulting in 2801 subcarriers within a 350-MHz measured BW. Because the TX (arbitrary waveform generator) and the RX (digital oscilloscope) are physically separated during the measurement, they are synchronized by two rubidium clocks disciplined by GPS satellites for accurate delay estimation. 

\begin{figure}[!t]
    \centering
    \includegraphics[height=\linewidth]{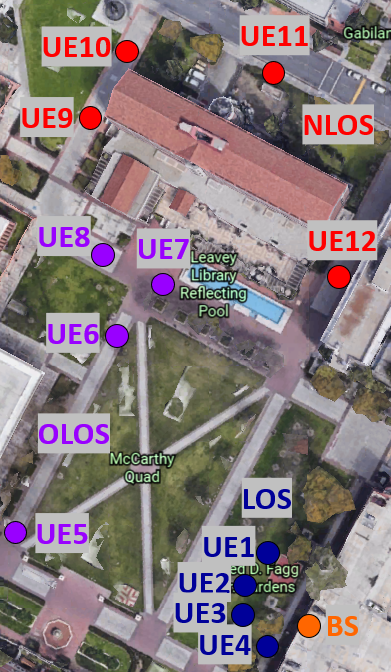}
    \caption{Map of the outdoor measurement campaigns. UEs at LOS locations are marked blue, UEs with trees blocking the LOS path are marked purple (obstructed LOS), and UEs with paths blocked by building are marked red (NLOS).}
    \label{fig:outdoor}
\end{figure}

The outdoor measurements were conducted at the northeast side of the University of Southern California (USC) University Park Campus, see Fig. \ref{fig:outdoor}, where the BS was positioned on top of a four-story high parking structure. In total, there were 12 UE locations and three cases (four UEs per case). The first case is a line-of-sight (LOS) case, where UEs were positioned close to the parking structure with LOS paths available (marked in blue). The second case is an obstructed-line-of-sight (OLOS), where UEs were spread out through the quad farther away from the parking structure, with trees blocking the LOS path (marked in purple). The last case is a NLOS case, where UEs were surrounding the library with the LOS path blocked by building (marked in red). 

\subsection{Indoor Measurements with Virtual Array}
For the indoor measurements, we selected for the RX side a rotating horn antenna that forms a virtual massive MIMO array (Fig. \ref{virtual_array}). Although indoor channel characteristics can also be measured using the real-time channel sounder with switched array, we opted for a different measurement method to diversify the setups. Also, using a precision VNA as the TX and RX can provide more accurate estimates, albeit it is feasible only for short-distance measurements. We conducted each measurement twice at two frequency bands (2.4--2.5 GHz and 5--7 GHz). 

We altered several setup configurations for the indoor measurements, including the waveform, the horn antenna, and the number of antenna positions. 201 frequency points were used for the 2.4--2.5 GHz band (500 kHz frequency spacing), whereas 1601 frequency points were used for the 5--7 GHz band (1.25 MHz frequency spacing). Using a larger frequency spacing with respect to the outdoor setup is possible because we expect a smaller excess delay. The intermediate frequency (IF) BW of the VNA was set to 500 kHz. The coarse frequency spacing and the wide IF BW helped to reduce the measurement time of the VNA-based channel sounder. The total measurement time was 75 seconds for each 2.4--2.5 and 5--7 GHz measurements per position. 

We also use two different horn antennas (both with 20 dBi gain) in the indoor measurements for the two frequency bands. The 3 dB beamwidth of the horn antenna used at 2.4 GHz is 12 degrees, whereas that for the 5--7 GHz band is 15--19 degrees, varying across the frequency. Therefore, we sample the azimuth every 12 degrees at 2.4 GHz, resulting in 30 azimuth points per elevation; and every 15 degrees at 5--7 GHz, resulting in 24 azimuth points per elevation. There are three elevation angles in both measurement setups; the antenna faces straight horizontally, at 10 degrees down from the horizontal plane, and at 10 degrees up from the horizontal plane. Therefore, the total antenna positions per location are $30 \times 3 =90$ in the 2.4--2.5 GHz band and $24 \times 3 =72$ in the 5--7 GHz band measurements.

\begin{figure}[!t]
    \centering
    \includegraphics[width=\linewidth]{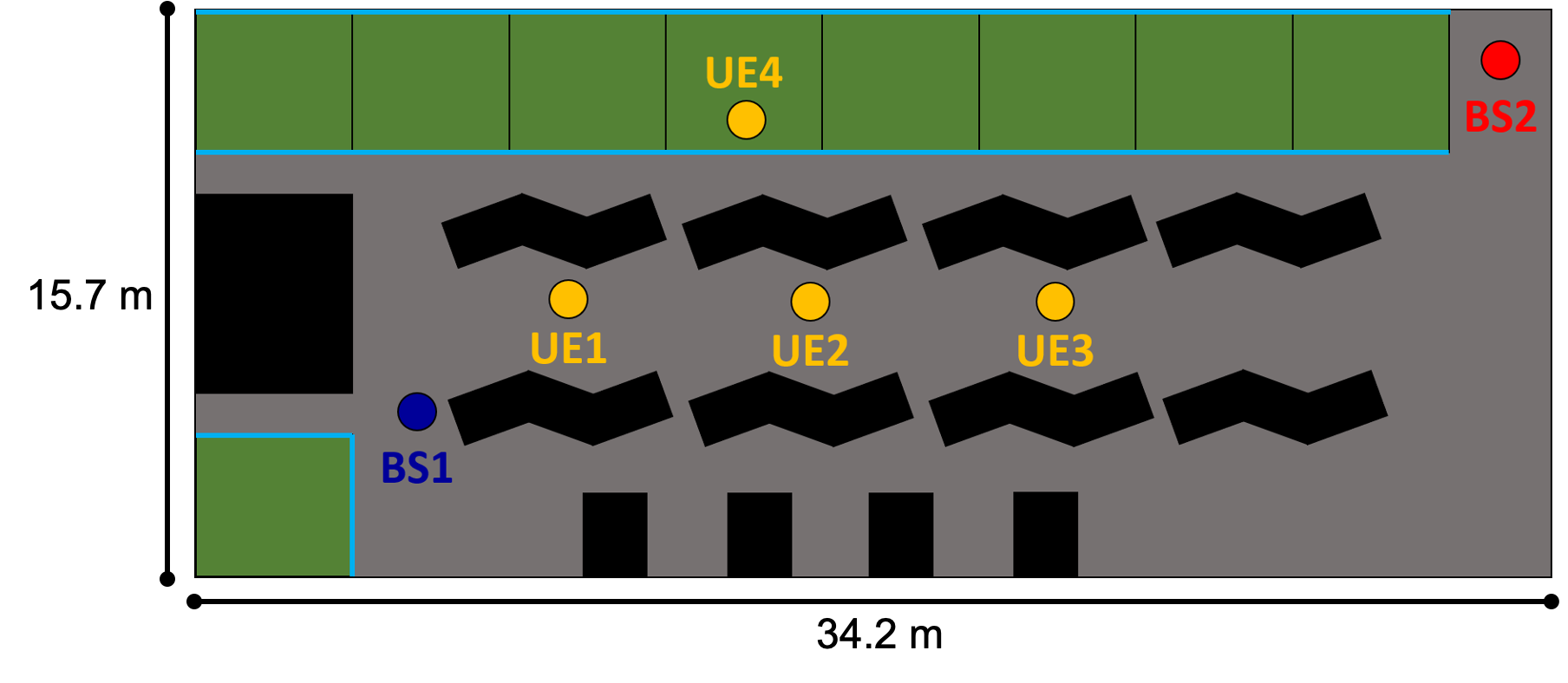}
    \caption{Map of the indoor measurement campaigns: The location of the BS serving the UEs at LOS locations is marked blue, another location of the BS serving the same UEs from an NLOS location is marked red, and the four locations of the UEs are marked yellow. The tables are marked in black, and small rooms are marked in green. The sky blue edges indicate glass doors and windows.}
    \label{fig:Leavey_2nd}
\end{figure}

The indoor measurements were conducted on the second floor of the Leavey Library, USC (Fig. \ref{fig:Leavey_2nd}). The virtual array was placed in two corners of the floor, whereas the omnidirectional antenna was moved to four different locations. The first BS (RX) location had LOS paths available from the UEs (TX), whereas the second BS location had NLOS paths from the UEs. The UE and the BS were positioned at 1.55 m height and 2.47 m height, respectively. The doors and windows of the rooms were made of glass. During the measurements, all Wi-Fi access points on the floor were covered with absorbers to prevent interference.


\section{Analysis of Results}
\subsection{3.325-3.675 GHz Outdoor}
We first analyze the 3.325--3.675 GHz outdoor measurements with switched array. We select 35 MHz around the 3.5 GHz center frequency (3.4825--3.5175 GHz) as a training band, while the remaining 315 MHz are used to validate the extrapolation results. The SAGE algorithm based on the VSS and the DOA models was applied to each location. A total of 10 paths are used for the VSS model, and 40 paths are used for the DOA model, where the number of paths increases until MSE $<$ -10 dB within the training band. LOS (UE1--UE4), OLOS (UE5--UE8), and NLOS (UE9--UE12) cases are analyzed separately.

\begin{figure}[!t]
    \centering
    \subfloat[VSS model]{\includegraphics[width=\linewidth]{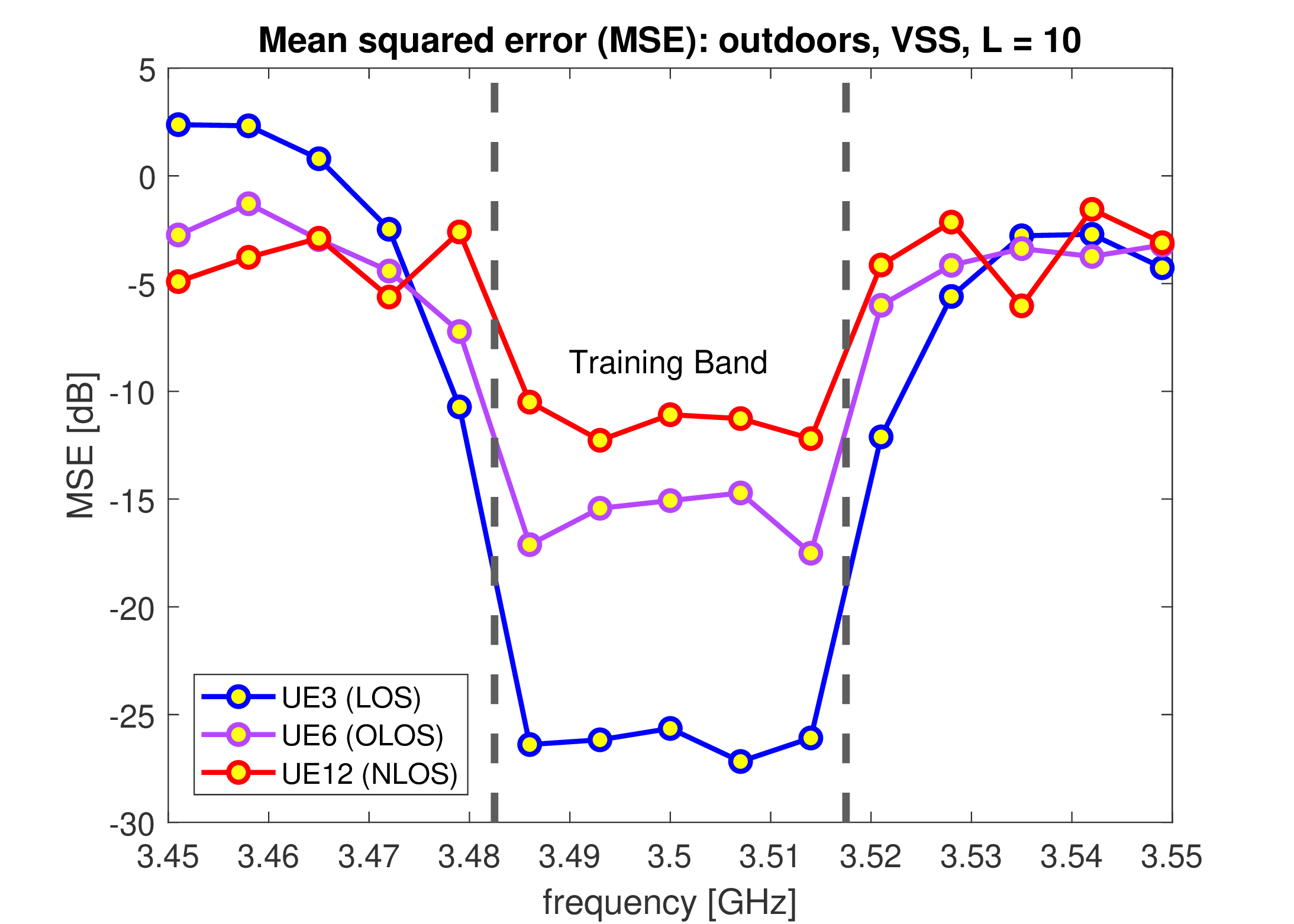}%
    \label{MSE3_vss}}
    \newline
    \subfloat[DOA model]{\includegraphics[width=\linewidth]{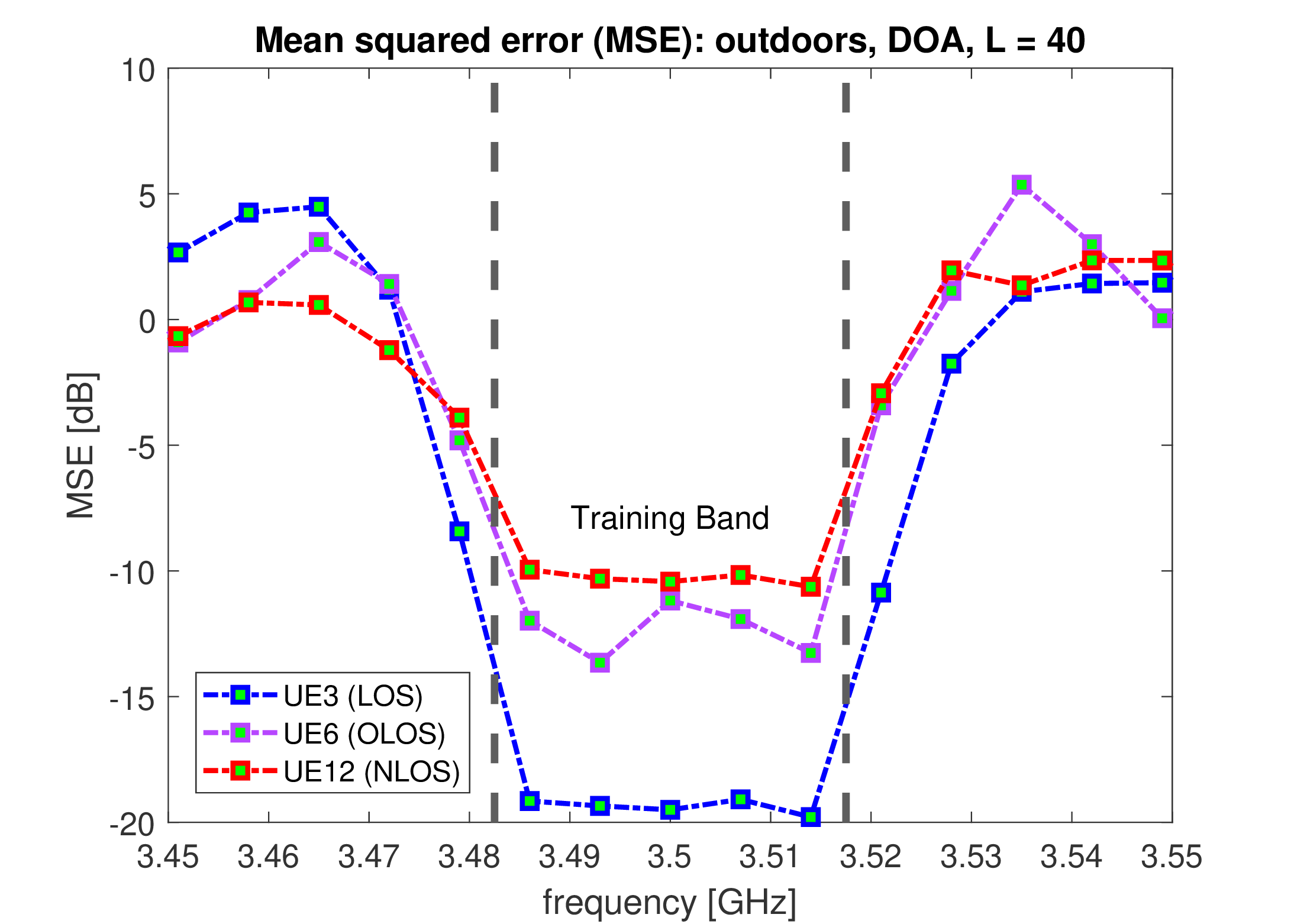}%
    \label{MSE3_DOA}}
    \caption{MSE of the estimated/extrapolated channel (outdoors, UE 3/6/12)}
\end{figure}

Fig. \ref{MSE3_vss} and Fig. \ref{MSE3_DOA} show the MSE. Fig. \ref{MSE3_vss} indicates the results based on the VSS model, whereas Fig. \ref{MSE3_vss} indicates those based on the DOA model. The blue lines evaluate UE3 (LOS); the purple lines, UE6 (OLOS); and the red lines, UE12 (NLOS). The sample UEs were chosen at random. The frequencies within the gray dashed lines indicate the training band, whereas the frequencies outside the band are the frequencies to be extrapolated. The graph clearly shows that within the training band, the SAGE algorithm based on all models at all locations estimates the channel reasonably well. The LOS shows the most accurate estimate, followed by the OLOS, and finally the NLOS. Among the VSS and the DOA models, the VSS estimates the channel more accurately even with smaller number of paths because---as explained in Sec. \ref{model}---the VSS model estimates more parameter values than the DOA model does ($2ML+L$ vs $5L$) in order to improve its fitness to the measured channel data within the training band. Unfortunately, the MSE quickly deteriorates outside the training band in all cases, which implies that the extrapolated channel deviates from the measured channel.

\begin{figure*}[!t]
    \centering
    \subfloat[Averaged BE (VSS)]{\includegraphics[width=0.5\linewidth]{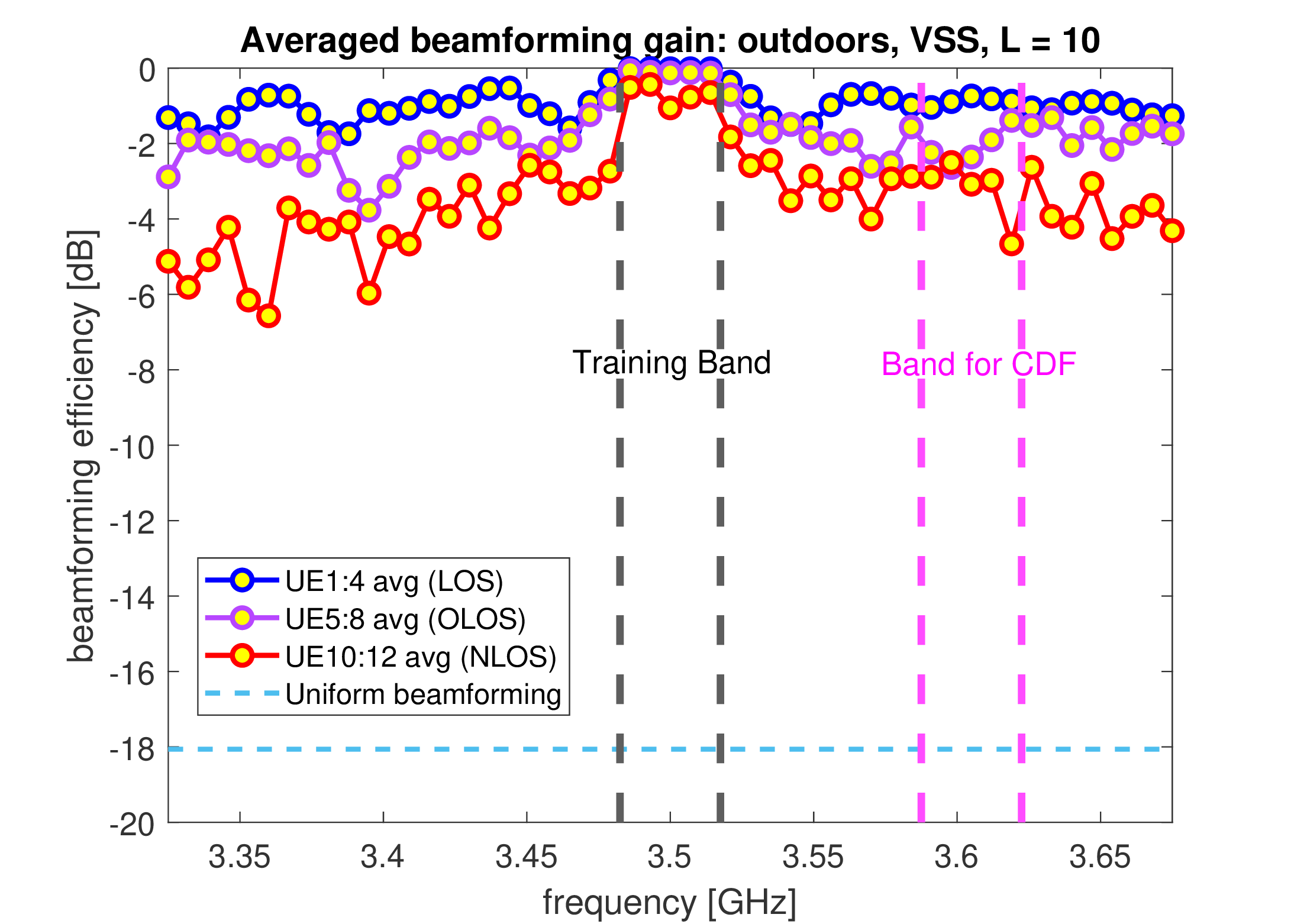}%
    \label{RBG_VSS}}
    \subfloat[CDF of 12 UEs with varying L (VSS)]{\includegraphics[width=0.5\linewidth]{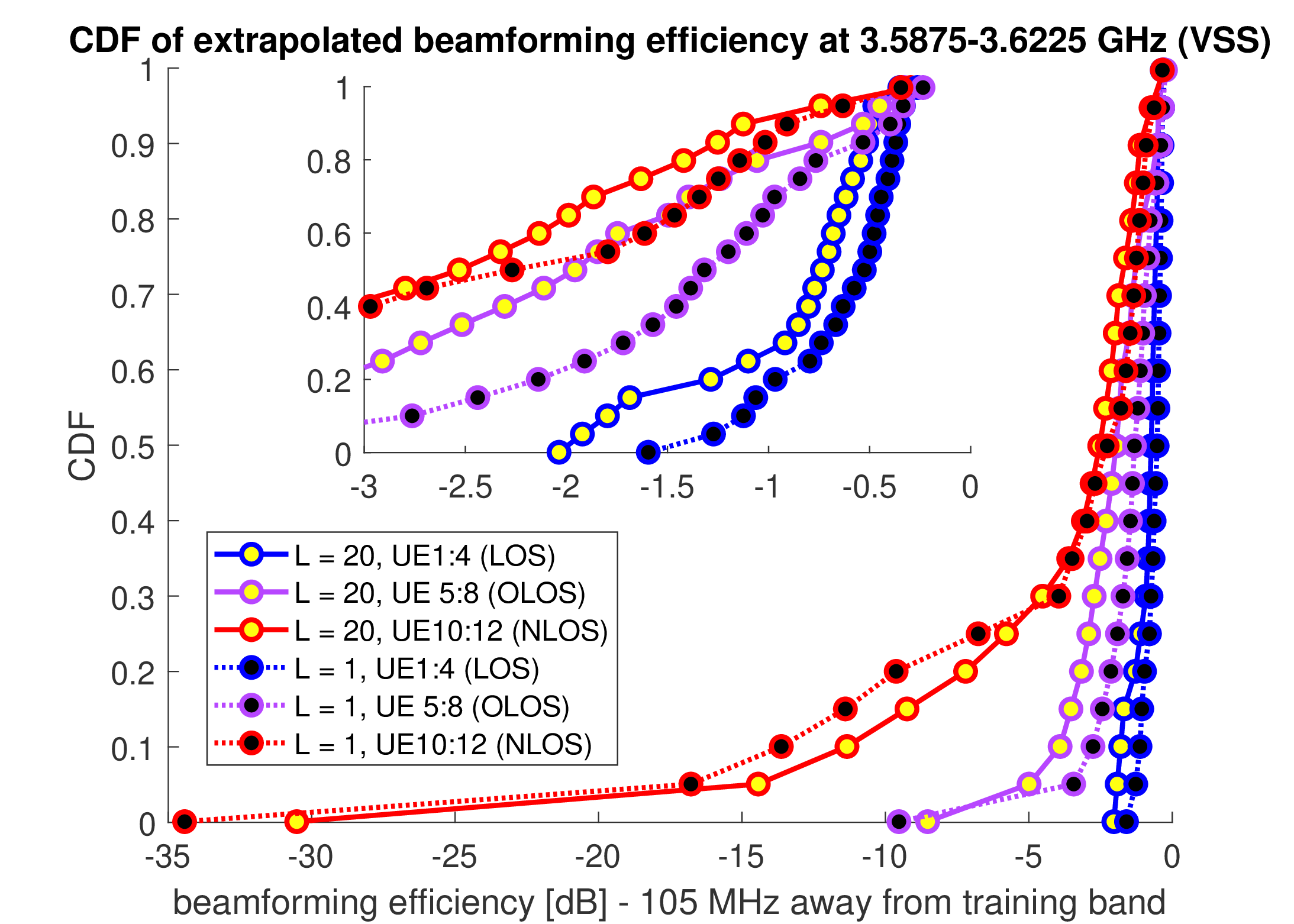}%
    \label{CDF_VSS}}
    \hfil
    \subfloat[Averaged BE (DOA)]{\includegraphics[width=0.5\linewidth]{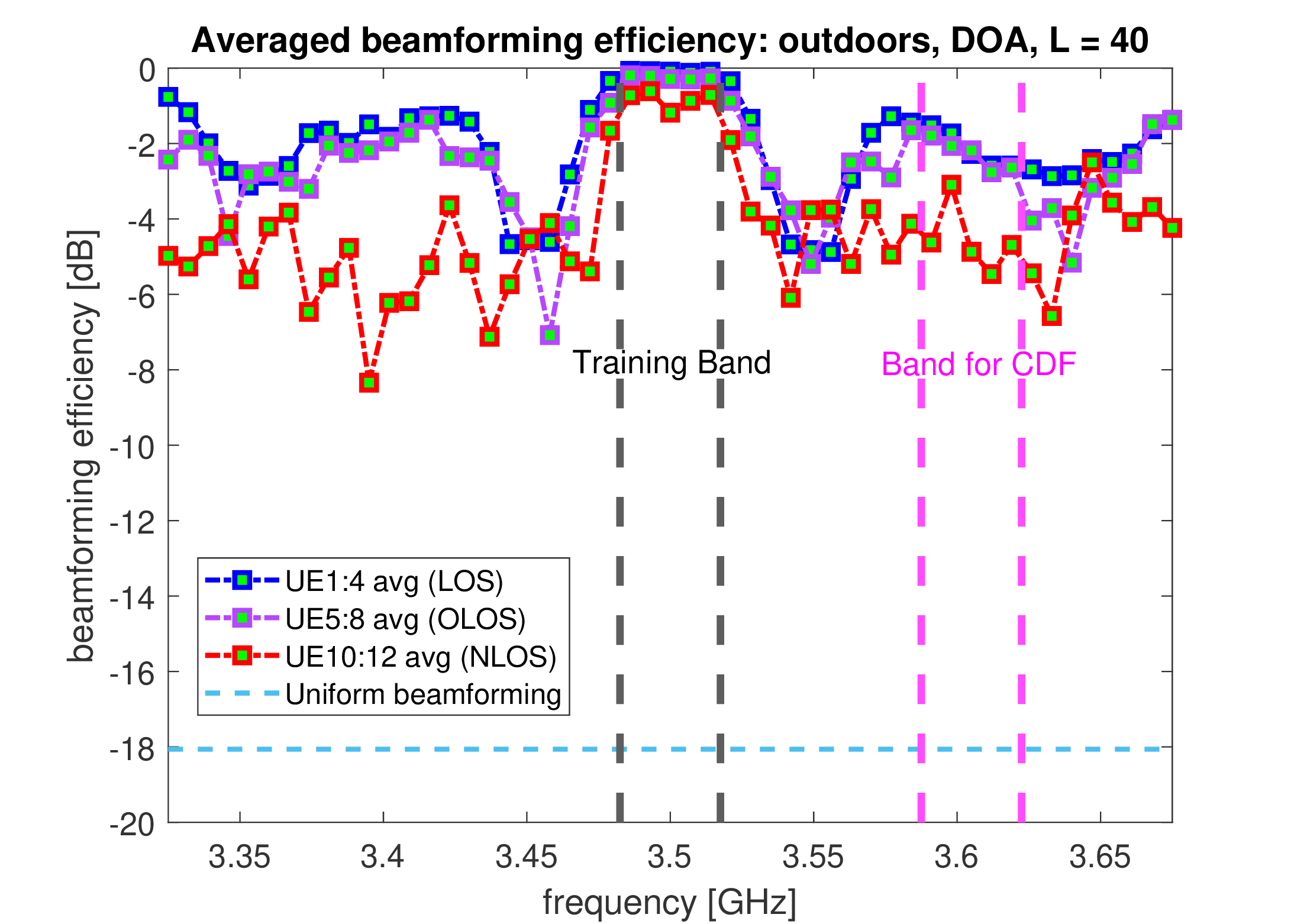}%
    \label{RBG_DOA}}
    \subfloat[CDF of 12 UEs with varying L (DOA)]{\includegraphics[width=0.5\linewidth]{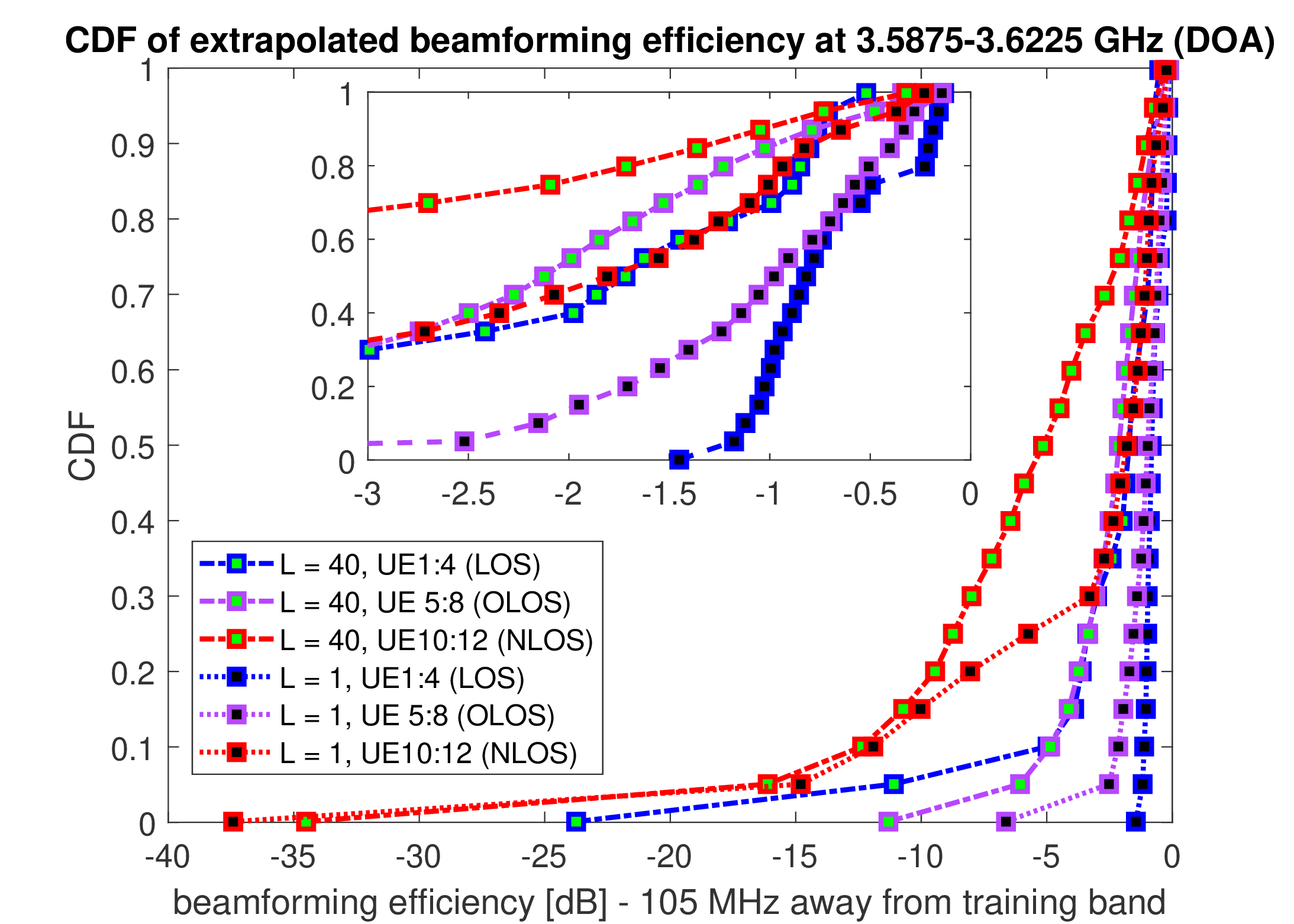}%
    \label{CDF_DOA}}
    \caption{Averaged BE over frequency and BE CDF (not averaged) 105 MHz away from the training band (outdoors, UE 1:12)}
    \label{BG3}
\end{figure*}

However, the MSE is not the most practical figure of merit when using channel extrapolation for communication system design purposes. Even if the extrapolated channel is not exactly similar to the measured channel, a communication system may still perform almost as well as if it has the true (measured) channel available. For example, the BE can be robust to errors in the channel estimate. Indeed, in practice, the so-called user specific reference signals are embedded in the stream to each user so that they can correct for a potential phase mismatch. This effect is not taken into account by the MSE, but is well-considered by the BE. The results of the BE in Fig. \ref{BG3} indeed provide alternative views on channel extrapolation. The BEs are averaged per case (4 UEs) using the VSS (Fig. \ref{RBG_VSS}) and the DOA models (Fig. \ref{RBG_DOA}). Also, the BE cumulative distributive functions (CDFs) in all 12 locations (the raw values in all UEs, not the averaged values), 105 MHz away from the training band (marked in pink in Fig. \ref{RBG_VSS} and \ref{RBG_DOA}), are shown in Fig. \ref{CDF_VSS} and Fig. \ref{CDF_DOA}, with additional results when $L=1$.

Several points need to be made here. First, in general, the DOA model provides higher variance and more deviation from the ground-truth CSI than the VSS model does. This may be attributed to the following factors: 1) the imperfect calibration of the antenna array and 2) insufficient parameters for estimating each MPC. Therefore, the VSS model is more practical to use than the DOA model for extrapolation because the former does not require any calibration data. Moreover, the computing speed is quicker in the former than in the latter. In other words, the VSS model has three main advantages: higher accuracy/performance, no need for array calibration and reduced implementation complexity.

Second, regardless of the model, using only one path is usually better (Fig. \ref{CDF_VSS} and Fig. \ref{CDF_DOA}). The outdoor measurements, even in the NLOS case, are better explained when using a single path rather than when multiple paths are used. Therefore, choosing many paths can result in overfitting during the extrapolation.

Lastly, although every case outperforms uniform beamforming, the performance deteriorates in the order of the LOS case, the OLOS case, then the NLOS case. In Fig. \ref{CDF_VSS}, the LOS performs well, with all BE values greater than -1.5 dB when $L=1$. In the OLOS case, more than 90 percent of the BE values are greater than -3 dB. In contrast, only 60 percent of the BE was greater than -3 dB in the NLOS case. Overall, in terms of BE, HRPE-based extrapolation shows potential, especially in the LOS cases in which a single path dominates the channel.

\begin{figure}[!t]
    \centering
    \subfloat[SE for UE4 (LOS)]{\includegraphics[width=\linewidth]{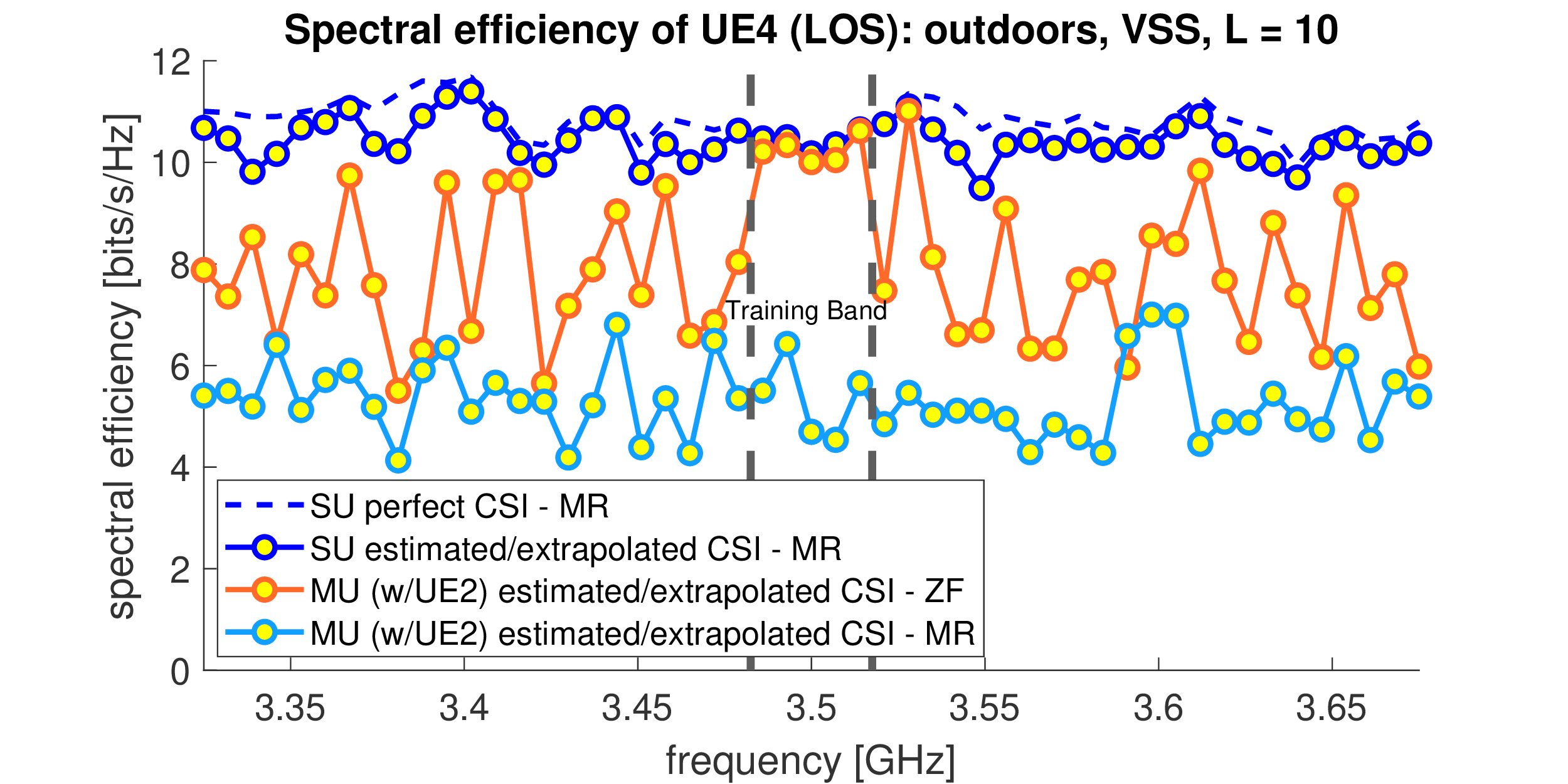}%
    \label{SE3_LOS}}
    \hfil
    \subfloat[SE for UE5 (OLOS)]{\includegraphics[width=\linewidth]{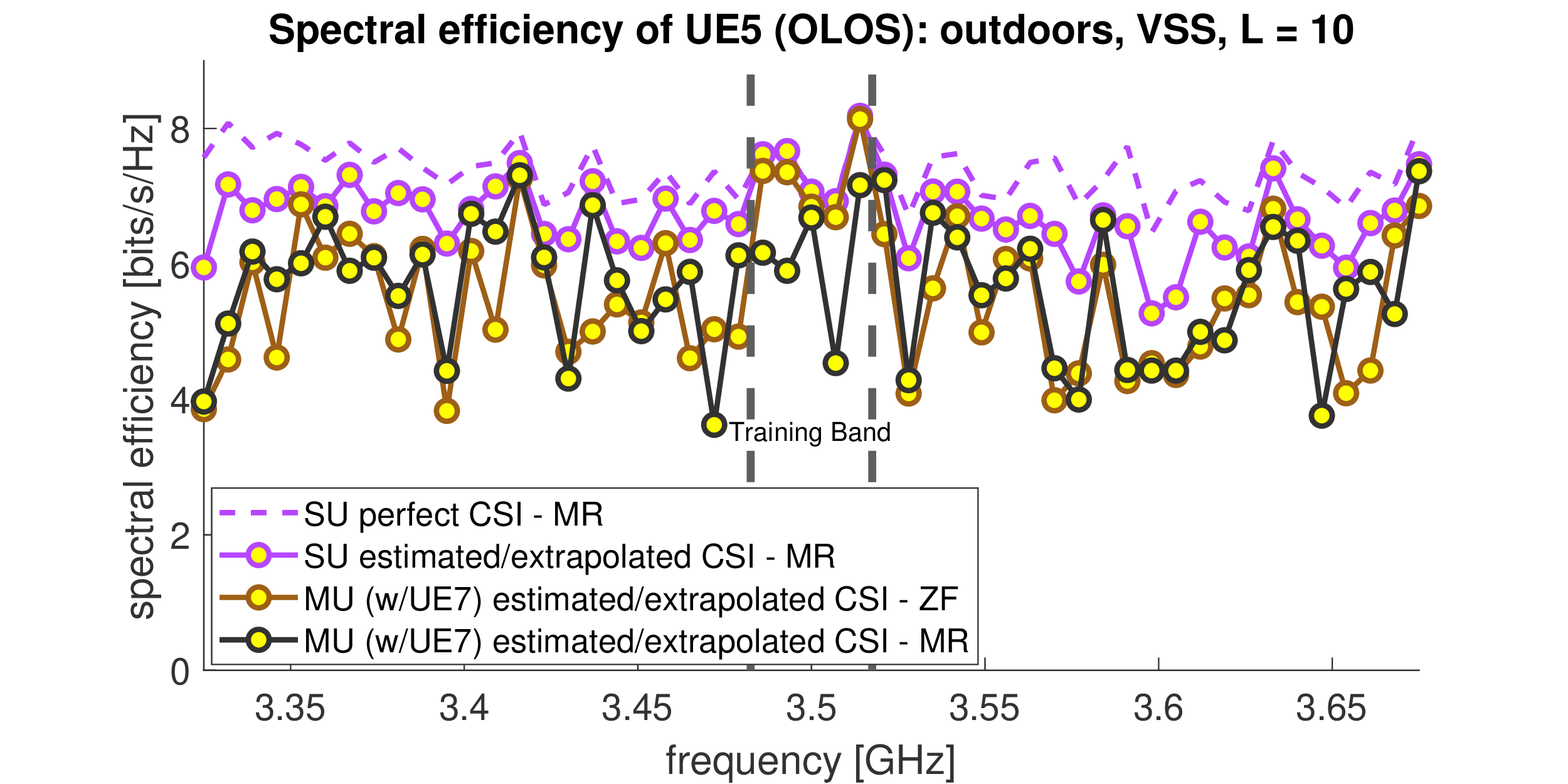}%
    \label{SE3_OLOS}}
    \hfil
    \subfloat[SE for UE12 (NLOS)]{\includegraphics[width=\linewidth]{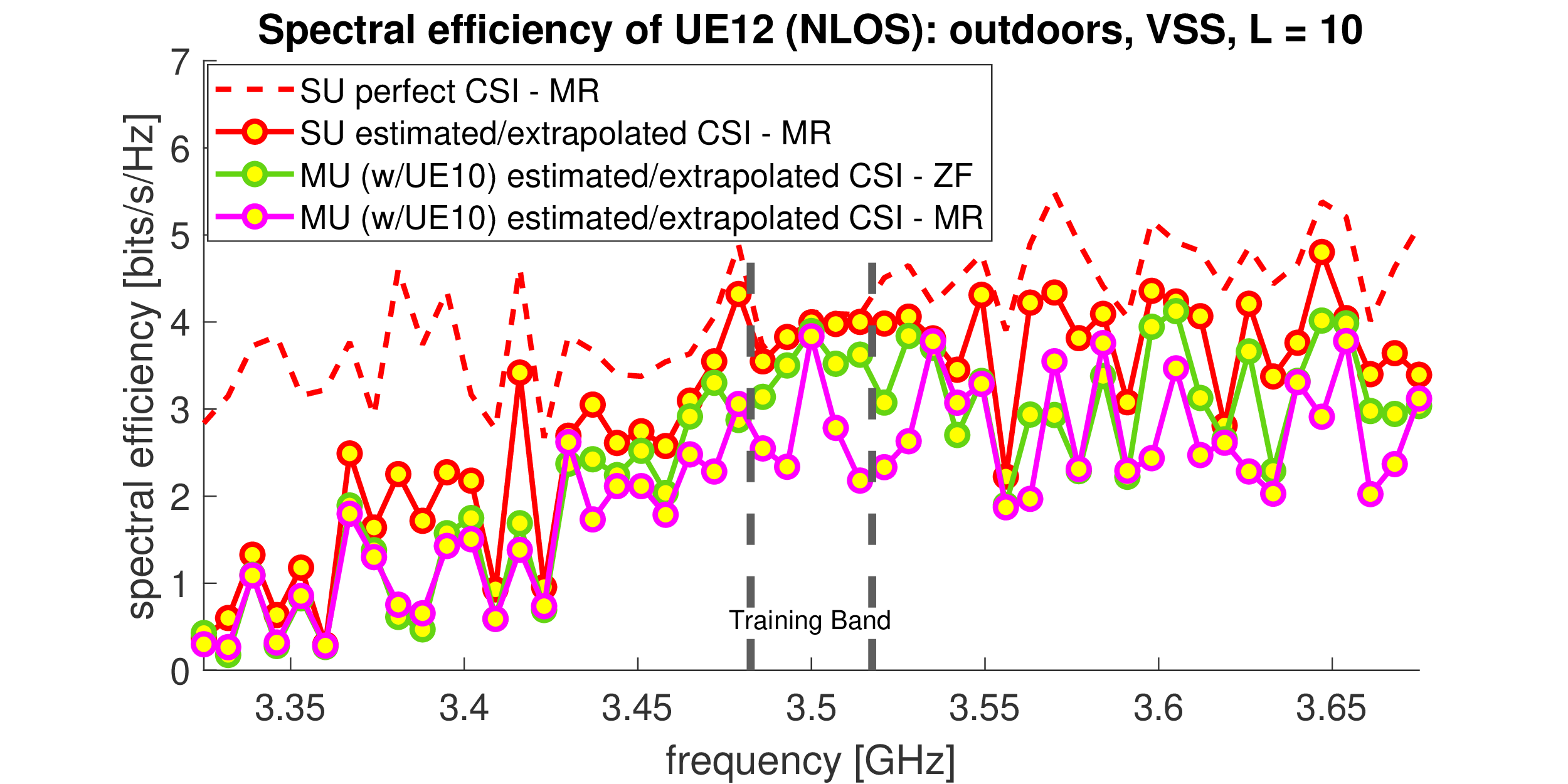}%
    \label{SE3_NLOS}}
    \caption{Spectral efficiency comparing single-user/multi-user scenarios with the estimated channels based on the VSS model (outdoors, UE 4/5/12)}
    \label{SE3}
\end{figure}

Fig. \ref{SE3} shows the spectral efficiency plots of both single-user and multiuser systems, which are achieved through the VSS model using 10 paths.\footnote{The DOA model is omitted here due to lack of space and since it has similar characteristics to the VSS model.} For the transmit SNR toward UE $n$, $\frac{{\sigma_{s^{(n)}}}^{2}}{{\sigma_{w^{(n)}}}^{2}} = 10^{10}$ (100 dB) is used assuming 30 dBm transmit power from outdoor BS and -70 dBm noise detected by UE. In the multiuser scenarios, 2 UEs within the same case (UE2$\&$4/5$\&$7/10$\&$12) are served together, where 2 UEs are chosen randomly from each LOS/OLOS/NLOS case. Accordingly, the three figures indicate the performances of UE4/5/12. Each figure contains four plots. There are single-user performance with ground-truth CSI (which is also very close to the multiuser ZF performance with ground-truth CSI---not shown in the figure), single-user performance with estimated/extrapolated CSI, multiuser ZF performance with estimated/extrapolated CSI, and multiuser MR performance with estimated/extrapolated CSI.

The results show that the extrapolated single-user spectral efficiency matches closely with the single-user spectral efficiency based on the ground-truth CSI. The deviations become larger as we move to the OLOS and then to the NLOS cases. Also, despite using only two UEs, the multiuser scenarios significantly deviate from the single-user scenario. This indicates that interference exists at the extrapolated frequencies even for the well-separated UEs, albeit such interference does not exist within the training band. 

Lastly, the ZF performs better than the MR in a LOS scenario, where the noise power is relatively smaller than the interference power. Meanwhile, they perform somewhat similarly in OLOS/NLOS scenarios, where the noise power dominates the SINR value due to the decreased signal strength. Overall, in terms of spectral efficiency, extrapolation works best for single-user scenario and when the UE is in LOS.

\subsection{2.4--2.5 and 5--7 GHz Indoor}
Next, the 2.4--2.5 and 5--7 GHz indoor measurements are analyzed together. We select 20 MHz between 2.4 and 2.42 GHz to serve as a training band for the first frequency band, and then select 100 MHz between 5.0 and 5.1 GHz to be a training band for the second frequency band. Different variations of the frequency bands are also selected to observe the effects of the training band size. Again, we apply the SAGE algorithm twice to the eight BS/UE combinations (2 BS locations and 4 UE locations) in accordance with the VSS and the DOA models. A total of 10 and 60 paths ($L$) are used, respectively.

\begin{figure}[!t]
    \centering
    \subfloat[2.4-GHz case]{\includegraphics[width=\linewidth]{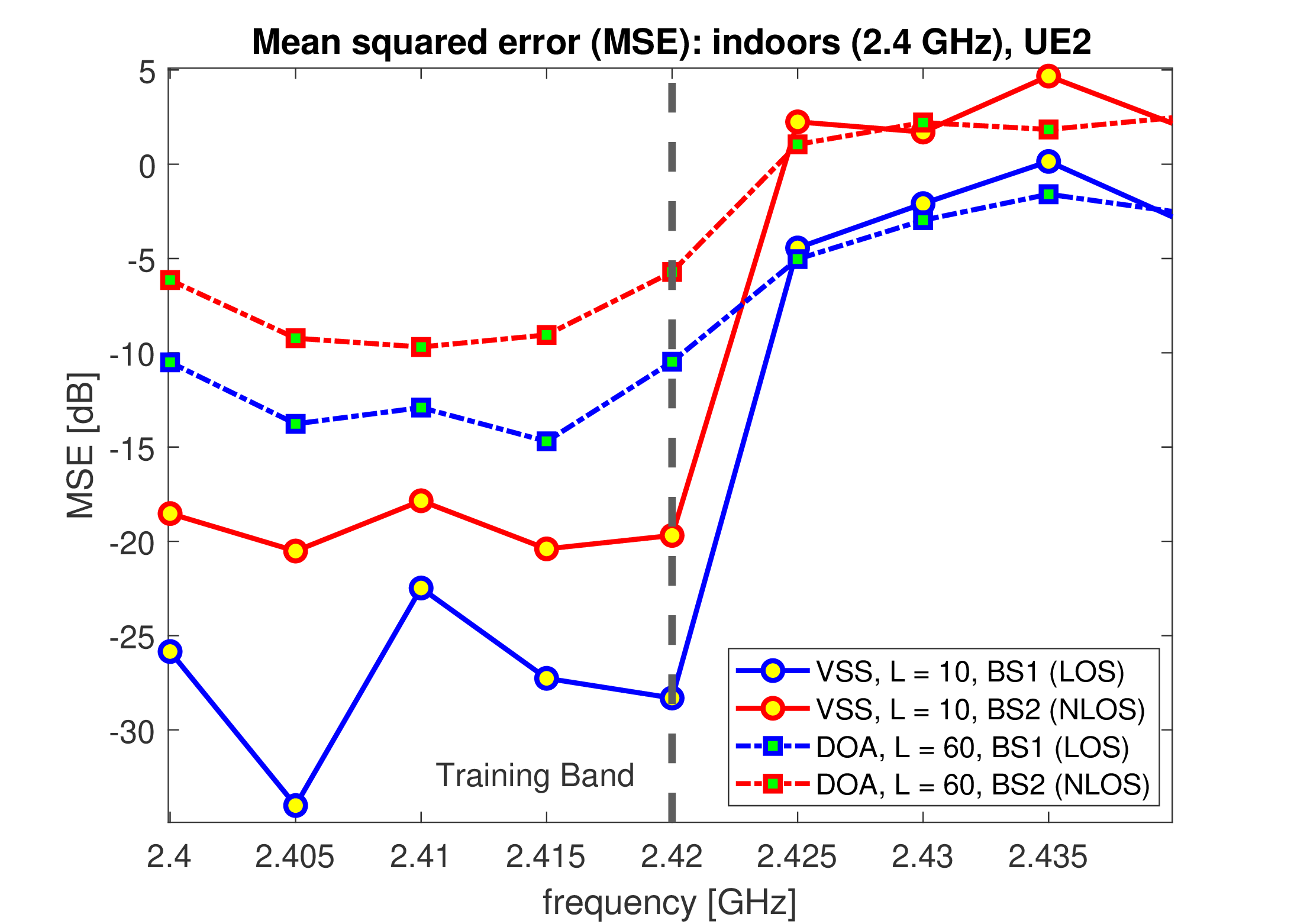}%
    \label{MSE2}}
    \newline
    \subfloat[5-GHz case]{\includegraphics[width=\linewidth]{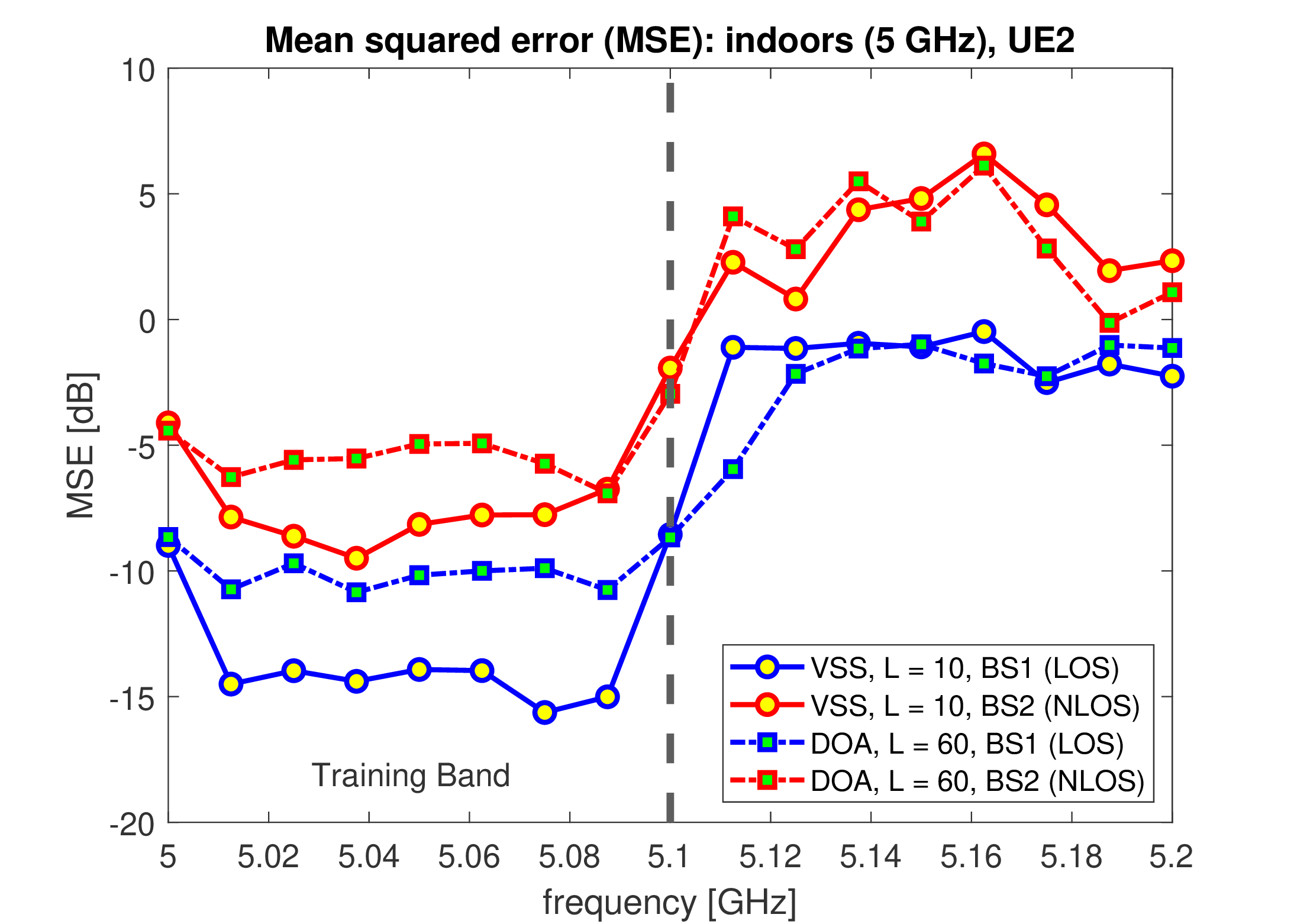}%
    \label{MSE5}}
    \caption{MSE of estimated/extrapolated channel (indoors, UE 2)}
    \label{MSE25}
\end{figure}

In Fig. \ref{MSE25}, the plots to the left of the gray dashed lines show the MSE of the estimated channels in the training bands. UE2 is arbitrarily selected, in which it is assumed to be served by BS1 (LOS) or BS2 (NLOS). The MSE is below -10 dB in the LOS case with the frequencies within the training band. This implies that the parameter values estimated by the SAGE algorithm can explain the channel accurately using the two channel models provided in Sec. \ref{model}. 

However, the NLOS case provides a relatively higher MSE, especially in the 5 GHz case. This is mainly because choosing the most accurate parameters for each MPC becomes difficult, especially when the frequency band widens and when the LOS path does not exist. For the frequencies outside the training band, the MSE performance degrades quickly. Similar to the 3.5 GHz outdoor case, the MPC parameter values change quickly when moving to another frequency. 

\begin{figure*}[!t]
    \centering
    \subfloat[Averaged BE (2.4--2.5 GHz)]{\includegraphics[width=0.5\linewidth]{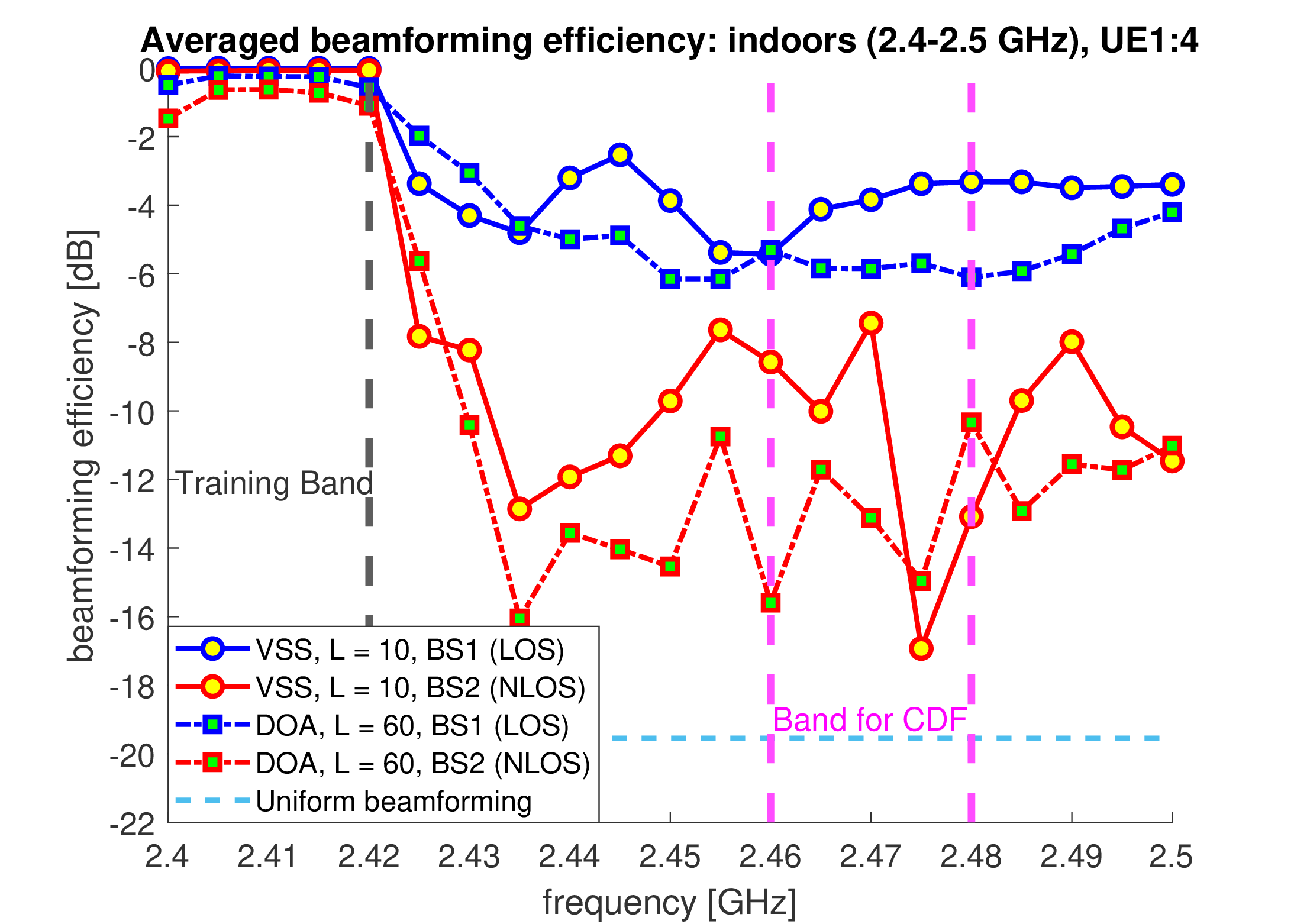}%
    \label{BG2}}
    \subfloat[CDF of 4 UEs with varying L (2.46--2.48 GHz)]{\includegraphics[width=0.5\linewidth]{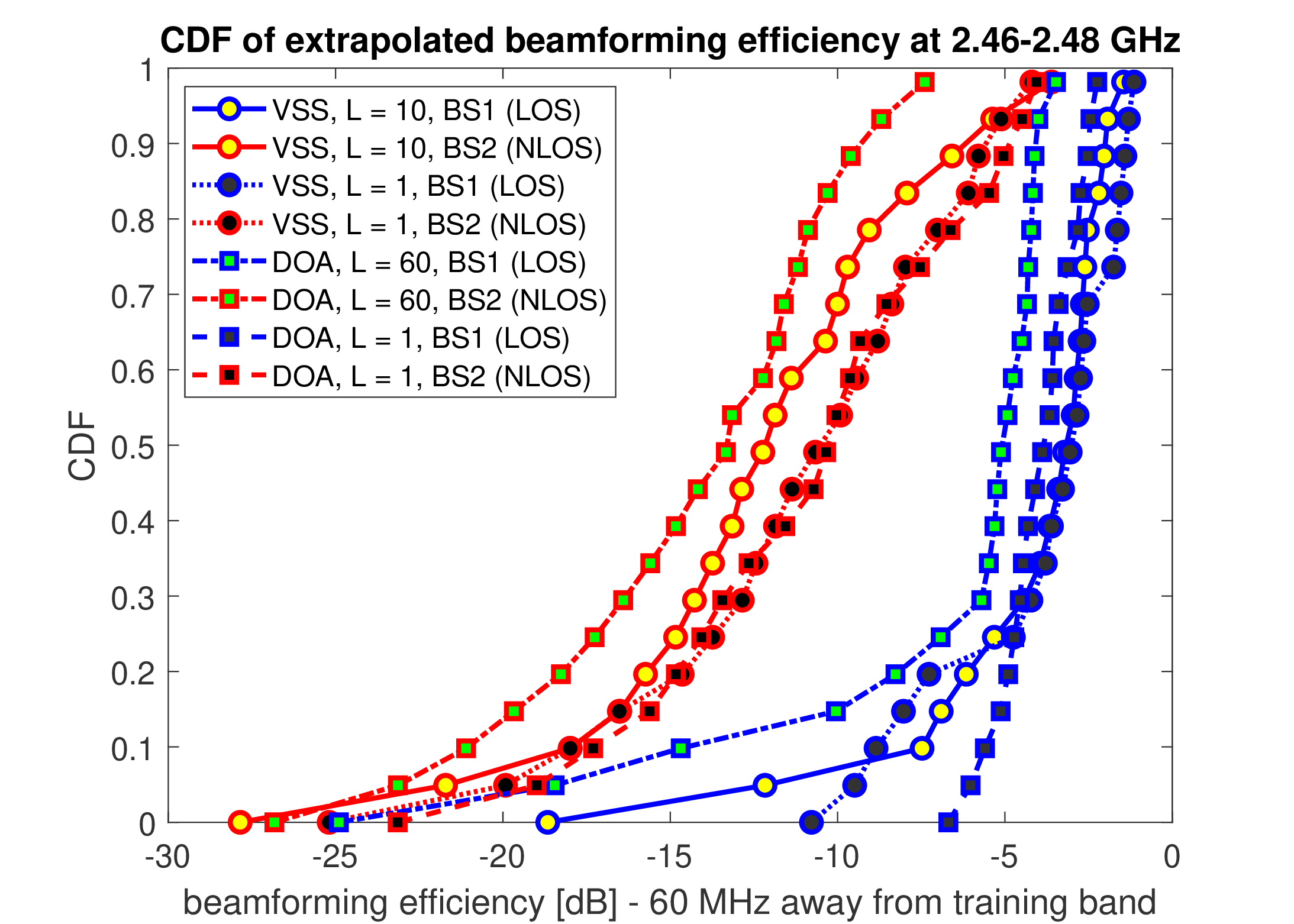}%
    \label{CDF2}}
    \hfil
    \subfloat[Averaged BE (5--7 GHz)]{\includegraphics[width=0.5\linewidth]{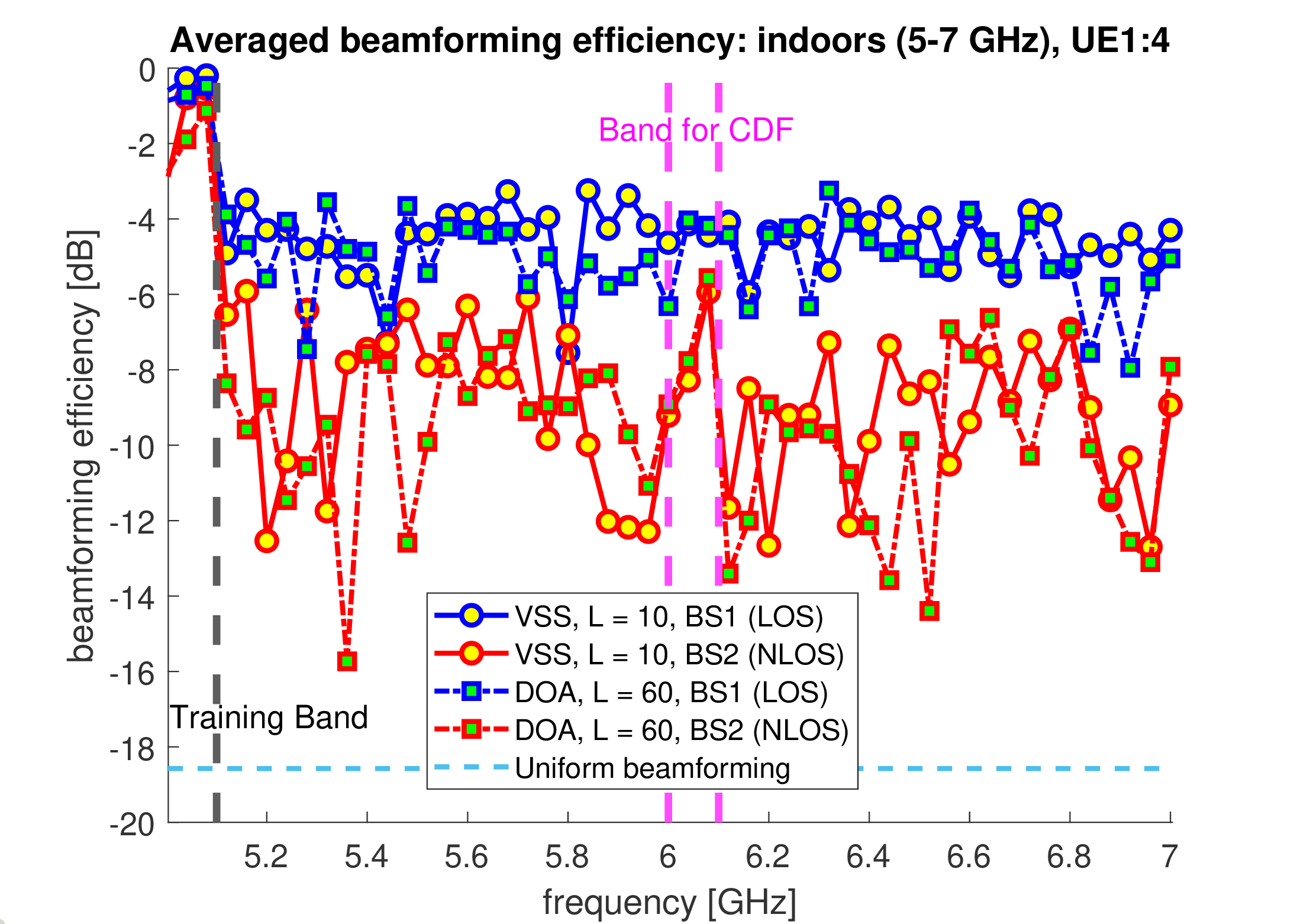}%
    \label{BG5}}
    \subfloat[CDF of 4 UEs with varying L (6.0--6.1 GHz)]{\includegraphics[width=0.5\linewidth]{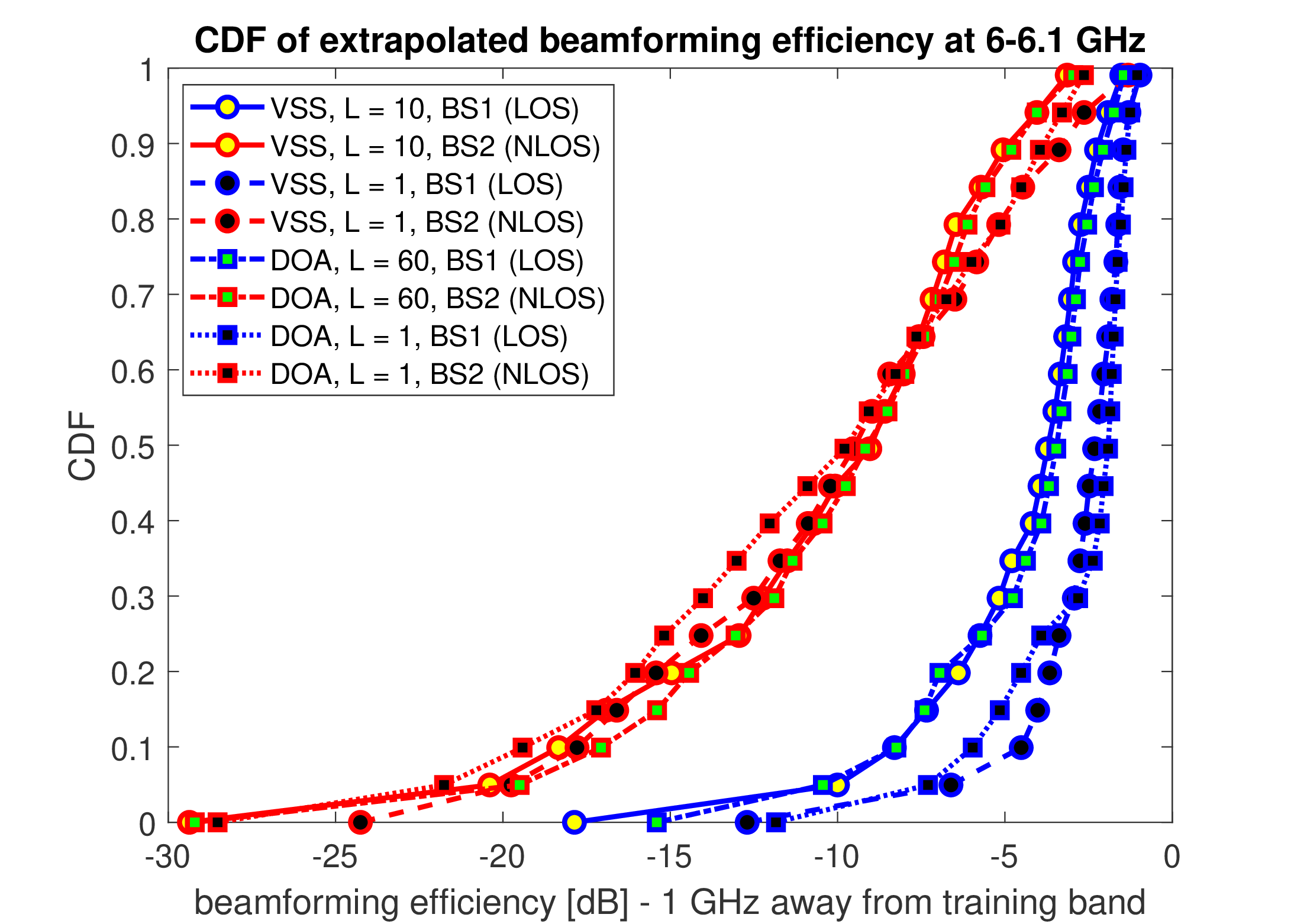}%
    \label{CDF5}}
    \caption{Averaged BE over frequency and BE CDF (not averaged) 60/1000 MHz away from the training band (indoors, UE 1:4)}
    \label{BG25}
\end{figure*}

Again, the MSE performance seems to discourage the use of channel extrapolation based on the SAGE algorithm. Yet, the BE in Fig. \ref{BG25} shows that channel extrapolation could be useful in the LOS cases. The BE in Fig. \ref{BG2} and \ref{BG5} is close to 0 dB in all LOS cases within the training band. This indicates that the BGs achieved through the estimated channel response and through the ground truth channel response are very similar. In the NLOS cases within the training band, although the VSS model in the 2.4 GHz case provides very little beamforming loss, the other cases result in BE ranging from -1 to -3 dB, which can still be acceptable.

The BE outside the training band performs worse when indoors than when outdoors. Unlike in Fig. \ref{CDF_VSS} and in \ref{CDF_DOA}, which show that most extrapolated channels provide -3 dB or greater BE at 105 MHz away from the training band, less than 50 percent of the LOS cases in Fig. \ref{CDF2} and less than 70 percent in Fig. \ref{CDF5} provide greater than -3 dB BE. The figures also show that one path ($L=1$) provide higher BE. This might be due to difficulty of extrapolating the many other paths than the LOS path that contribute to the channel in the indoors environment. The fact that the large extrapolation range (1.9 GHz) in the 5 GHz case and the beamforming performance do not have clear correlation show that the LOS path can be extrapolated easily. 

In the NLOS cases, the BE is generally much worse than that in the LOS cases, thereby discouraging the use of extrapolation. Again, the results show that it is difficult to extrapolate without a dominant LOS path and when the extrapolation is within a rich scattering environment, as theoretically foreseen in \cite{rottenberg2020performance}. In terms of the model, the VSS model performs better than the DOA model in the 2.4--2.5 GHz case, whereas both models perform similarly in the 5--7 GHz case. 

\begin{figure}[!t]
    \centering
    \subfloat[SE for UE3 (LOS)]{\includegraphics[width=\linewidth]{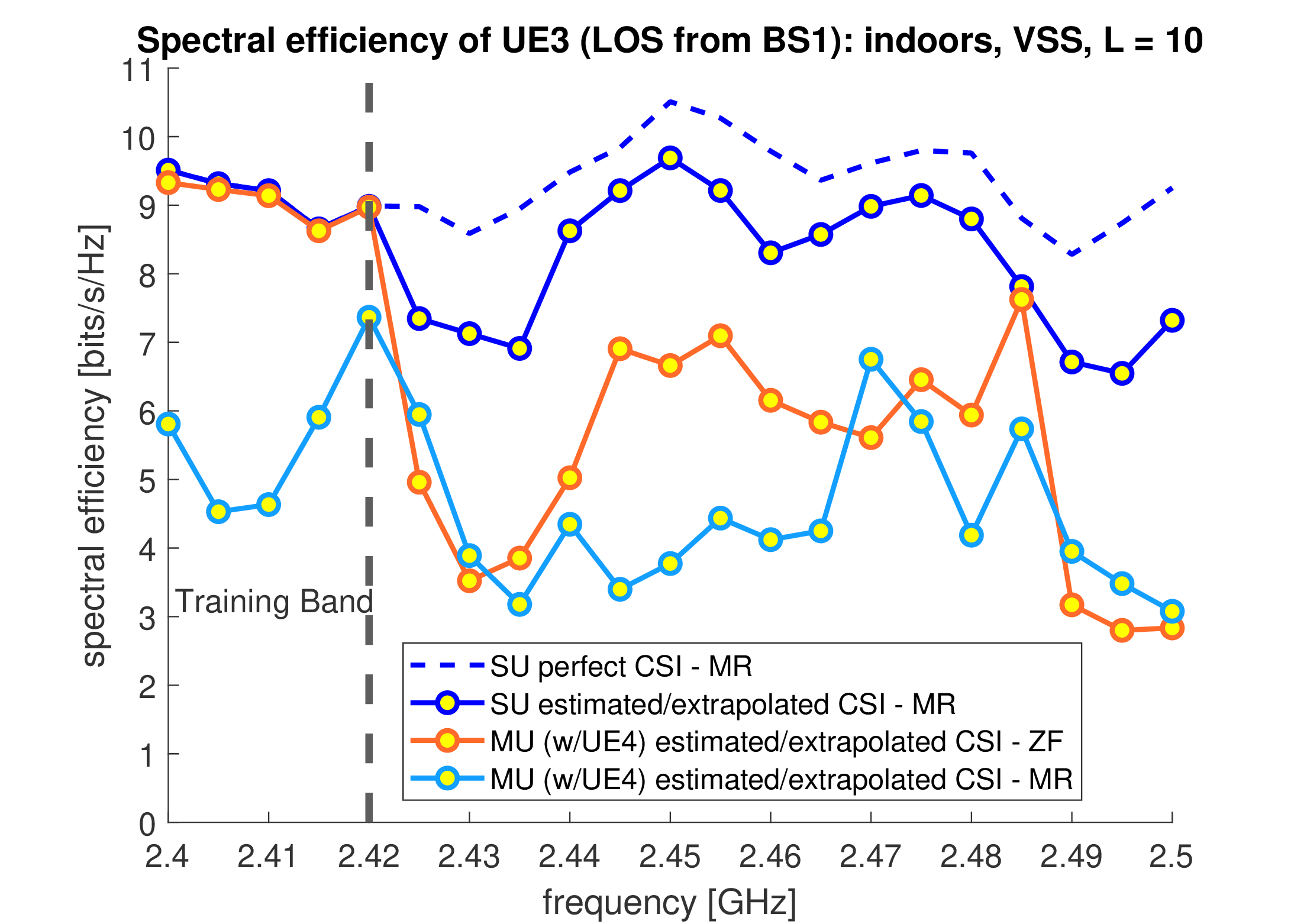}%
    \label{SE2L}}
    \newline
    \subfloat[SE for UE3 (NLOS)]{\includegraphics[width=\linewidth]{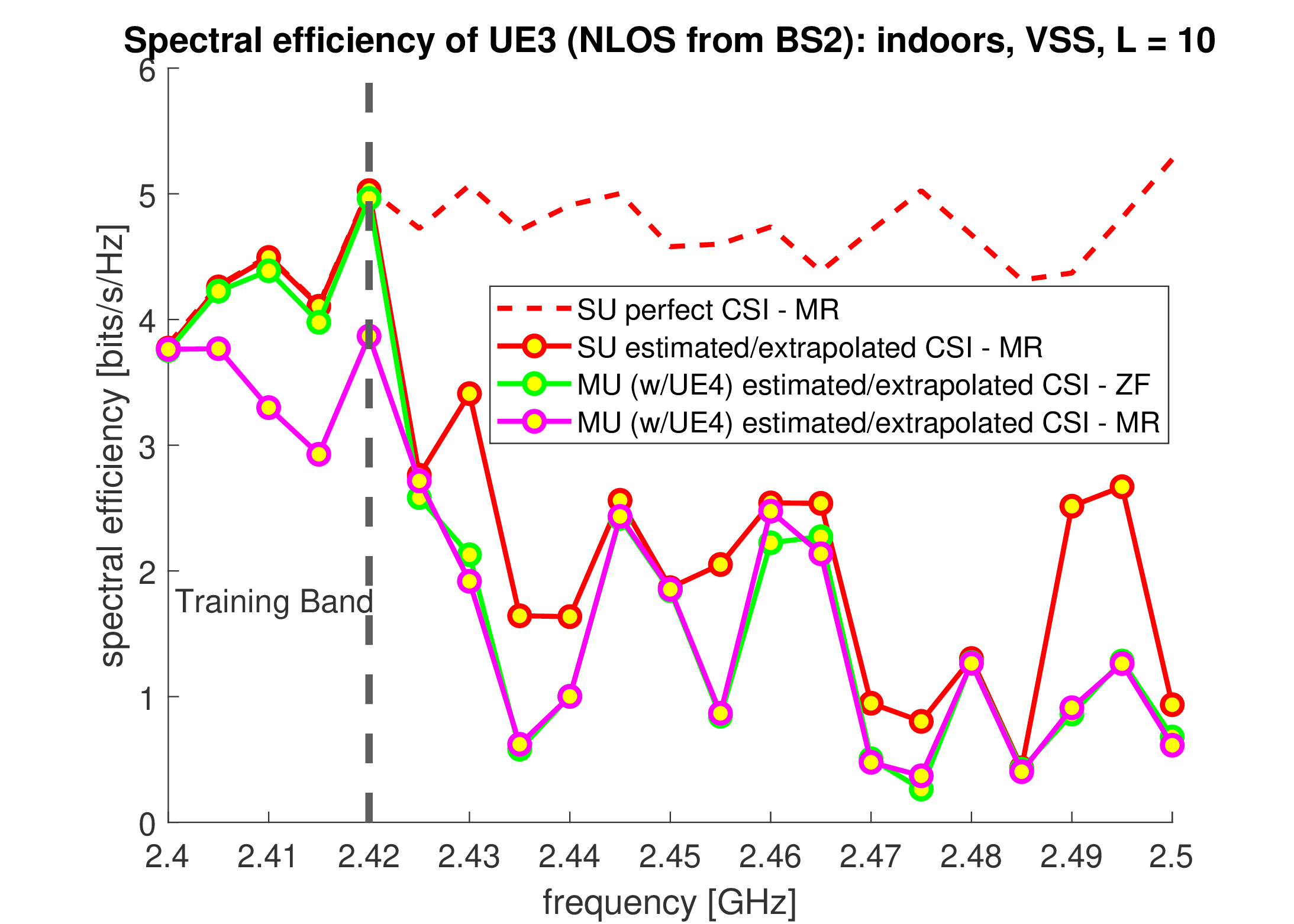}%
    \label{SE2N}}
    \caption{Spectral efficiency comparing the single-user/multiuser scenarios with the estimated channels based on the VSS model (indoors, UE 3)}
    \label{SE2}
\end{figure}
 
Lastly, Fig. \ref{SE2} shows the spectral efficiency of UE3 served by BS1 (LOS) and BS2 (NLOS).\footnote{Only the results of the 2.4--2.5 GHz case using the VSS model is plotted due to space restriction and to their similarity to the DOA results.} There are two scenarios: when only the UE3 is served (single-user) and when both UE3 and UE4 are served simultaneously (multiuser). Two different precodings (i.e., MR and ZF) are used in the multiuser case. The transmit SNR toward UE $n$ was set to $\frac{{\sigma_{s^{(n)}}}^{2}}{{\sigma_{w^{(n)}}}^{2}} = 10^{8}$ (80 dB) assuming 10 dBm transmit power from indoor BS and -70 dBm noise dectected by UE.

The results show that although the single-user performance in the LOS case using the extrapolated channels has small performance difference from the perfect CSI (differing by about 2 bits/s/Hz at maximum), it does not perform as well as in the LOS case under the outdoor environment (as in Fig. \ref{SE3_LOS}). As expected, the extrapolated channel provides large loss in spectral efficiency in the NLOS case. Specifically, the single-user performance and the multiuser performance in the NLOS case are rather similar in the extrapolated bands. Although the ZF provides better than the MR in the LOS case, they are overlapping in the NLOS case. Overall, extrapolation is difficult in multiuser scenarios, especially in the NLOS case, whereas it performs relatively better in the single-user LOS scenario. 

% if have a single appendix:
%\appendix[Proof of the Zonklar Equations]
% or
%\appendix  % for no appendix heading
% do not use \section anymore after \appendix, only \section*
% is possibly needed

% use appendices with more than one appendix
% then use \section to start each appendix
% you must declare a \section before using any
% \subsection or using \label (\appendices by itself
% starts a section numbered zero.)
%

\section{Conclusion}

This paper presents the results of our experimental study that determined the feasibility of channel extrapolation using the MPC parameters at a selected training frequency band, with the main goal of eliminating large overheads in FDD massive MIMO systems. The MPC parameters at the training frequency band are attained through the HRPE based on the VSS and the DOA models. The empirical data obtained from three massive MIMO channel measurement setups under two settings (i.e., outdoor and indoor environments) are used for validations. 

%To eliminate large overheads in FDD massive MIMO systems, an experimental study to determine the feasibility of channel extrapolation using the parameters of MPCs at a selected training frequency band attained through the HRPE that estimated the channel frequency response at a desired frequency band was presented. The empirical data obtained by three massive MIMO channel measurement setups in outdoor and indoor environments were used to verify our results.
%

The results show that the best extrapolation performance can be achieved when 1) a channel contains a dominant LOS path, 2) no interacting objects surround the BS (outdoor), and 3) the UEs are well separated. One of the reasons why it is difficult to extrapolate in the NLOS cases is the unpredictability of small-scale fading. Another source of error may come from the simplicity of the channel model, which assumes that the MPCs are planar waves with perfect vertical polarization. Such model assumption will not hold when the BS is too close to the interacting objects. Therefore, the suggested extrapolation method will perform best in the FDD massive MIMO systems that follow conditions 1)--3). 

We conjecture that a particularly appealing case for future research could be the channels in stationary unmanned aerial vehicle (UAV) communication systems. UAV communication systems usually involve high LOS probability, function in outdoor environment, have well-separated aerial vehicles, and have sufficiently long channel coherence time. Also, the distributed massive MIMO systems with high probability of LOS paths between a subset of BS antennas and the UEs may benefit from channel extrapolation. Meanwhile, the FDD massive MIMO systems utilizing concentrated array for terrestrial applications, which mostly involve NLOS cases, may still require at least partial DL pilot and feedback overheads or better extrapolation techniques in order to improve performance. 

Another notable result is that the simpler VSS model-based extrapolation, which estimates an abstract antenna array pattern without array calibration and with lower implementation complexity, performs better than the DOA model-based extrapolation, which uses measured calibration data of antenna array to find the DOAs of the MPCs. This is somewhat counterintuitive: having more information about the antenna array used during the measurement should provide better results; however, the opposite occurs. One possible explanation is the sensitivity of the antenna array to calibration and model errors. If the goal of the HRPE is to reconstruct an estimated channel based on observation rather than finding angular characteristics of the MPCs, then the VSS might be a preferable algorithm.

Several topics arising from the current work will be considered in the future, including: 1) the dependence of the extrapolation performance on the number and/or geometry (planar or cylindrical) of antennas at the BS and/or at the UEs, 2) extension and/or improvement of the channel model, including the polarization parameter and/or spherical wave model, and 3) the channel measurements of UAV-based massive MIMO systems or distributed massive MIMO systems to verify the HRPE-based channel extrapolation techniques for a realistic massive MIMO system dominated by the LOS path.

\appendices 

\section{SAGE Algorithm} \label{appendix}
The SAGE algorithm \cite{fleury1999channel} is a well-known parameter estimation algorithm that extends the expectation maximization (EM) algorithm by identifying the ``estimated parameters describing MPCs in a channel'', $\vect{\hat{\Psi}}$. Thus, it maximizes the likelihood of the ``observed complex channel frequency response'', $H(m, f_k)$, over all antennas (with index $m$) and frequency points (with index $k$):

\begin{align*}
    \vect{\hat{\Psi}} =\mathrm{arg}\min_{\hat{\vect{\Psi}}} \sum_{m}\sum_{k} | H(m,f_k) - \hat{H}(m, f_k;\hat{\vect{\Psi}})|^2.
\end{align*}

$\hat{H}(m, f_k;\hat{\vect{\Psi}})$ is the ``estimated complex channel frequency response'' constructed by the maximum likelihood estimated MPC parameters $\hat{\vect{\Psi}} = [\hat{\vect{\psi}}_1, \hdots ,\hat{\vect{\psi}}_L]$ and the VSS or the DOA models previously introduced in sec. \ref{model}.

This optimization problem is challenging because of 1) the high-dimensionality scaling with $L$ and 2) the nonlinear dependence on the path parameters. The SAGE algorithm provides an efficient suboptimal solution to the problem of relying on an iterative approach.
% Does SAGE approaches the problem 2? If so, in what way?

\begin{figure*} [!t]
    \centering
    \includegraphics[width=0.75\linewidth]{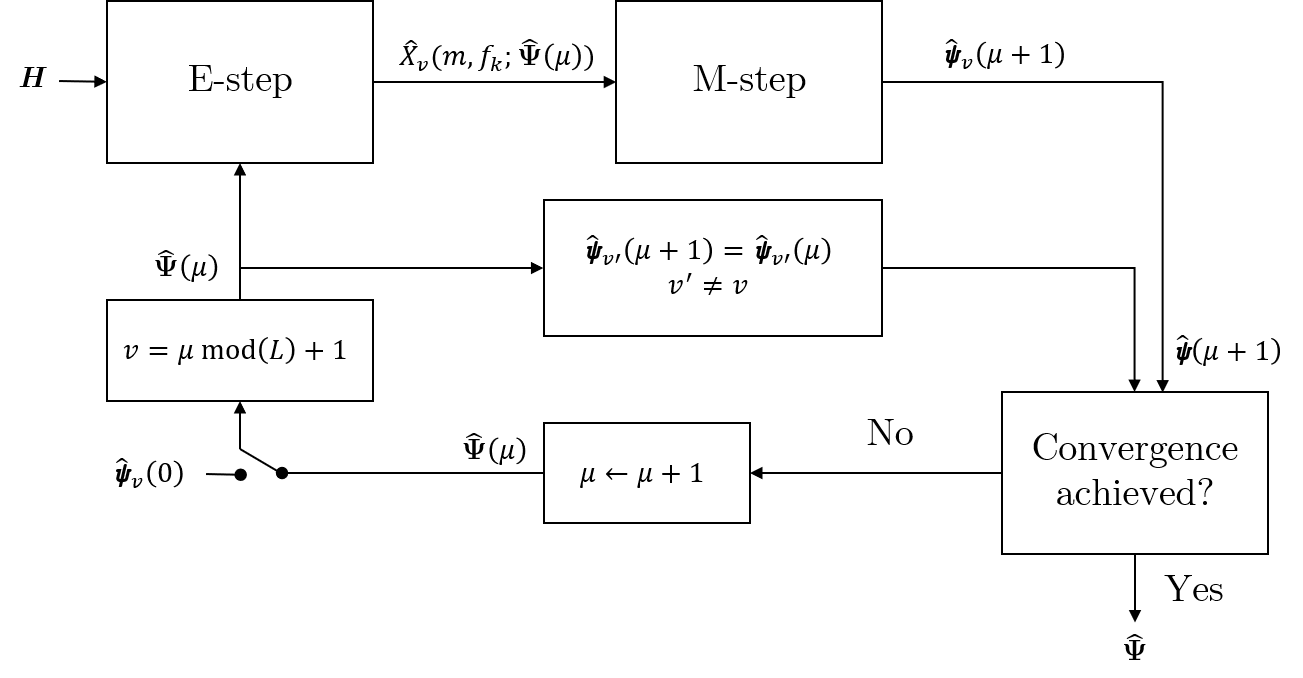}
    \caption{SAGE algorithm flow graph for the VSS model - $v$ can be replaced by $d$ for the DOA model}
    \label{fig:algorithm}
\end{figure*}

\begin{figure*} [!t]
    \centering
    \includegraphics[width=0.75\linewidth]{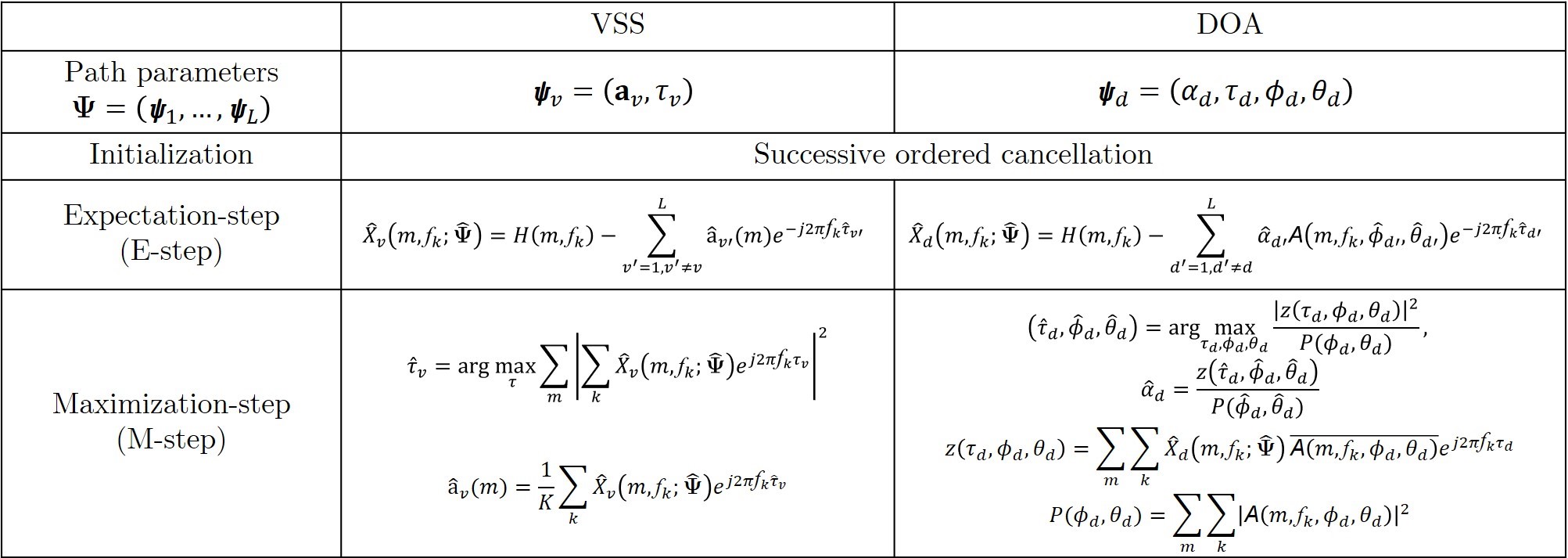}
    \caption{Descriptions of the SAGE algorithm steps for the VSS and DOA models}
    \label{fig:application}
\end{figure*}

Fig.~\ref{fig:algorithm} is the SAGE algorithm flow chart. The estimates of ``the ground truth parameters'', ${\vect{\Psi}}$, at iteration $\mu$, are denoted by $\hat{\vect{\Psi}}(\mu)$. The successive ordered cancellation is used to estimate the initial parameters, $\hat{\vect{\Psi}}(0)$, as explained in \cite{fleury1999channel}. At each iteration, only the parameters corresponding to one path, \textit{e.g.}, $\hat{\vect{\psi}}_v$ \textit{or} $\hat{\vect{\psi}}_d$, are optimized while the parameters for the other paths keep their past value ($v=\mu~\mathrm{mod}(L)+1$). Optimizing the parameters per path reduces the search dimensions by a factor $L$. A set of $L$ iterations, called an iteration cycle, is required to update the parameters for each of the $L$ paths. After an iteration cycle, each path is reestimated based on the updated values of other paths. 

Each iteration consists of two steps: 1) the expectation step and 2) the maximization step. The specific operations performed in the two steps in the VSS and DOA algorithms are shown in Fig.~\ref{fig:application}. In the expectation step, the interference due to the other paths is canceled from the measured channel response based on their current estimate. Then, during the maximization step, the parameters $\hat{\vect{\psi}}_v$ or $\hat{\vect{\psi}}_d$ are reestimated. To further reduce the complexity of the problem, the optimization as a function of the different parameters of each path is simplified into several 1-D searches over the predefined parameter grids. This optimizes each parameter at a time, thereby fixing all the other parameters except for the estimated parameter. The algorithm iterates until convergence or if the maximum number of iterations is achieved. The implementation is done through MATLAB software.

% use section* for acknowledgment
\section*{Acknowledgment}
The authors would like to thank A. Adame, A. Alvarado, Z. Cheng, Dr. C. U. Bas, Dr. N. Abbasi, and Dr. D. Burghal for their assistance in developing the channel sounder, in the measurement campaigns, and in the productive technical discussions.

% Can use something like this to put references on a page
% by themselves when using endfloat and the captionsoff option.
\ifCLASSOPTIONcaptionsoff
  \newpage
\fi

% trigger a \newpage just before the given reference
% number - used to balance the columns on the last page
% adjust value as needed - may need to be readjusted if
% the document is modified later
%\IEEEtriggeratref{8}
% The "triggered" command can be changed if desired:
%\IEEEtriggercmd{\enlargethispage{-5in}}

% references section

% can use a bibliography generated by BibTeX as a .bbl file
% BibTeX documentation can be easily obtained at:
% http://mirror.ctan.org/biblio/bibtex/contrib/doc/
% The IEEEtran BibTeX style support page is at:
% http://www.michaelshell.org/tex/ieeetran/bibtex/
%\bibliographystyle{IEEEtran}
% argument is your BibTeX string definitions and bibliography database(s)
%\bibliography{IEEEabrv,../bib/paper}
%
% <OR> manually copy in the resultant .bbl file
% set second argument of \begin to the number of references
% (used to reserve space for the reference number labels box)
%\begin{thebibliography}{1}

%\bibitem{IEEEhowto:kopka}
%H.~Kopka and P.~W. Daly, \emph{A Guide to \LaTeX}, 3rd~ed.\hskip 1em plus
%  0.5em minus 0.4em\relax Harlow, England: Addison-Wesley, 1999.

%\end{thebibliography}
\bibliography{mMIMO_extrapol}
\bibliographystyle{IEEEtran}

% biography section
% 
% If you have an EPS/PDF photo (graphicx package needed) extra braces are
% needed around the contents of the optional argument to biography to prevent
% the LaTeX parser from getting confused when it sees the complicated
% \includegraphics command within an optional argument. (You could create
% your own custom macro containing the \includegraphics command to make things
% simpler here.)
%\begin{IEEEbiography}[{\includegraphics[width=1in,height=1.25in,clip,keepaspectratio]{mshell}}]{Michael Shell}
% or if you just want to reserve a space for a photo:

%\begin{IEEEbiography}{Michael Shell}
%Biography text here.
%\end{IEEEbiography}

% if you will not have a photo at all:
%\begin{IEEEbiographynophoto}{John Doe}
%Biography text here.
%\end{IEEEbiographynophoto}

% insert where needed to balance the two columns on the last page with
% biographies
%\newpage

%\begin{IEEEbiographynophoto}{Jane Doe}
%Biography text here.
%\end{IEEEbiographynophoto}

% You can push biographies down or up by placing
% a \vfill before or after them. The appropriate
% use of \vfill depends on what kind of text is
% on the last page and whether or not the columns
% are being equalized.

%\vfill

% Can be used to pull up biographies so that the bottom of the last one
% is flush with the other column.
%\enlargethispage{-5in}

% that's all folks
\end{document}